\documentclass{aa}

\usepackage[varg]{txfonts}
\usepackage{natbib}
\usepackage{graphicx}
\usepackage{subcaption}
\usepackage{rotating}
\usepackage{multirow} 
\usepackage{xcolor} 
\usepackage{amsmath}
\usepackage{ulem} 

\usepackage{amssymb}
\newcommand{\msun}{\mbox{M$_{\odot}$\,}}

\newcommand{\zsol}{\mbox{Z$_{\odot}$\,}}
\newcommand{\zsun}{\mbox{Z$_{\odot}$\,}}

\newcommand{\kms}{\mbox{$\rm{km}\,s^{-1}$}}

\DeclareMathAlphabet{\mathsc}{OT1}{cmr}{m}{sc}
\def\testbx{bx}%
\DeclareRobustCommand{\ion}[2]{%
\relax\ifmmode
\ifx\testbx\f@series
{\mathbf{#1\,\mathsc{#2}}}\else
{\mathrm{#1\,\mathsc{#2}}}\fi
\else\textup{#1\,{\mdseries\textsc{#2}}}%
\fi}
\newcommand{\ha} {\mbox{H$\alpha$}\,}
\newcommand{\hb} {\mbox{H$\beta$}\,}

\newcommand{\Hb} {\mbox{H$\beta$}\,}

\newcommand{\Feii} {\ion{Fe}{ii}\,}

\newcommand{\Caii} {[\ion{Ca}{ii}]\,}

\newcommand{\Hii} {\ion{H}{ii}\,}
\newcommand{\Hei} {\ion{He}{i}\,}
\newcommand{\Heii} {\ion{He}{ii}\,}

\newcommand{\Nii} {[\ion{N}{ii}]\,}
\newcommand{\Oi} {[\ion{O}{i}]\,}

\newcommand{\Oii} {[\ion{O}{ii}]\,}
\newcommand{\Oiii} {[\ion{O}{iii}]\,}
\newcommand{\Sii} {[\ion{S}{ii}]\,}

\newcommand{\Neiii} {[\ion{Ne}{iii}]\,}

\newcommand{\Mgii} {\ion{Mg}{ii}\,}

\newcommand{\eg}[0]{$\textnormal{e.g. }$}
\newcommand{\ie}[0]{$\textnormal{i.e. }$}
\newcommand{\sub}[1]{_{\textnormal{\scriptsize{#1}}}}

\begin{document}

\title{The evolution of superluminous supernova LSQ14mo and its interacting host galaxy system\thanks{Based on observations at ESO, Program IDs: 191.D-0935, 094.D-0645, 096.A-9099} }

\author{T.-W.~Chen \inst{\ref{inst1},\ref{inst2}\thanks{E-mail: jchen@mpe.mpg.de}}
\and M.~Nicholl \inst{\ref{inst3}}
\and S. J.~Smartt \inst{\ref{inst4}}  
\and P. A.~Mazzali \inst{\ref{inst5},\ref{inst6}}
\and R. M.~Yates \inst{\ref{inst1}}
\and T. J.~Moriya \inst{\ref{inst7}}
\and C.~Inserra \inst{\ref{inst4}}  
\and N.~Langer \inst{\ref{inst2}} 
\and T.~Kr{\"u}hler \inst{\ref{inst1}}
\and Y.-C.~Pan \inst{\ref{inst8}}
\and R.~Kotak \inst{\ref{inst4}}
\and L.~Galbany \inst{\ref{inst9},\ref{inst10}}
\and P.~Schady \inst{\ref{inst1}}
\and P.~Wiseman \inst{\ref{inst1}}
\and J.~Greiner \inst{\ref{inst1}}
\and S.~Schulze \inst{\ref{inst11}}
\and A. W. S.~Man \inst{\ref{inst12}}
\and A.~Jerkstrand \inst{\ref{inst6}}
\and K. W.~Smith \inst{\ref{inst4}}
\and M.~Dennefeld \inst{\ref{inst13}}
\and C.~Baltay \inst{\ref{inst14}}
\and J.~Bolmer \inst{\ref{inst1},\ref{inst15}}
\and E.~Kankare \inst{\ref{inst4}}
\and F.~Knust \inst{\ref{inst1}}
\and K.~Maguire \inst{\ref{inst4}}
\and D.~Rabinowitz \inst{\ref{inst14}}
\and S.~Rostami \inst{\ref{inst14}}
\and M.~Sullivan \inst{\ref{inst16}}
\and D. R.~Young \inst{\ref{inst4}}
}

\institute{Max-Planck-Institut f{\"u}r Extraterrestrische Physik, Giessenbachstra\ss e 1, 85748, Garching, Germany\label{inst1}
\and Argelander Institute for Astronomy, University of Bonn, Auf dem H\"ugel 71, D-53121 Bonn, Germany\label{inst2}
\and Harvard-Smithsonian Center for Astrophysics, 60 Garden Street, Cambridge, Massachusetts 02138, USA\label{inst3}
\and Astrophysics Research Centre, School of Mathematics and Physics, Queen's University Belfast, Belfast BT7 1NN, UK\label{inst4}
\and Astrophysics Research Institute, Liverpool John Moores University, IC2, Liverpool Science Park, 146 Brownlow Hill, Liverpool L3 5RF, UK\label{inst5}
\and Max-Planck-Institut f{\"u}r Astrophysik, Karl-Schwarzschild-Str. 1, DE-85748 Garching-bei-M\"{u}nchen, Germany\label{inst6}
\and Division of Theoretical Astronomy, National Astronomical Observatory of Japan, National Institutes of Natural Sciences, 2-21-1 Osawa, Mitaka, Tokyo 181-8588, Japan\label{inst7}
\and Department of Astronomy and Astrophysics, University of California, Santa Cruz, CA 95064, USA\label{inst8}
\and Pittsburgh Particle Physics, Astrophysics, and Cosmology Center (PITT PACC)\label{inst9}
\and Physics and Astronomy Department, University of Pittsburgh, Pittsburgh, PA 15260, USA\label{inst10}
\and Department of Particle Physics and Astrophysics, Weizmann Institute of Science, Rehovot 76100, Israel\label{inst11}
\and ESO, Karl-Schwarzschild-Strasse 2, DE-85748 Garching-bei-M\"{u}nchen, Germany\label{inst12}
\and Institut d'Astrophysique de Paris, CNRS, and Universite Pierre et Marie Curie, 98 bis Boulevard Arago, F-75014 Paris, France\label{inst13}
\and Physics Department, Yale University, 217 Prospect Street, New Haven, CT 06511-8499, USA\label{inst14}
\and European Southern Observatory, Alonso de C{\'o}rdova 3107, Vitacura, Casilla 19001, Santiago 19, Chile\label{inst15}
\and Department of Physics and Astronomy, University of Southampton, Southampton, SO17 1BJ, UK\label{inst16}
}

\date{Received November 29, 2016 /
Accepted January 25, 2017 }

\abstract{
We present and analyse an extensive dataset of the superluminous supernova (SLSN) LSQ14mo ($z = 0.256$), consisting of a multi-colour lightcurve from $-30$\,d to +70\,d in the rest-frame (relative to maximum light) and a series of six spectra from PESSTO covering $-7$\,d to +50\,d. This is among the densest spectroscopic coverage, and best-constrained rising lightcurve, for a fast-declining hydrogen-poor SLSN. The bolometric lightcurve can be reproduced with a millisecond magnetar model with $\sim 4$\,\msun ejecta mass, and the temperature and velocity evolution is also suggestive of a magnetar as the power source.  
Spectral modelling indicates that the SN ejected $\sim 6$\,\msun of CO-rich material with a kinetic energy of $\sim 7 \times 10^{51}$\,erg, and suggests a partially thermalised additional source of luminosity between $-2$\,d and +22\,d. This may be due to interaction with a shell of material originating from pre-explosion mass loss.  
We further present a detailed analysis of the host galaxy system of LSQ14mo. PESSTO and GROND imaging show three spatially resolved bright regions, and we used the VLT and FORS2 to obtain a deep (five-hour exposure) spectra of the SN 
position and the three star-forming regions, which are at a similar redshift. 
The FORS spectrum at $+300$ days shows no trace of SN emission lines and 
we place limits on the strength of [O\,{\sc i}] from comparisons with other Ic supernovae. The deep spectra provides a unique chance to investigate spatial variations in the host star-formation activity and metallicity.
The specific star-formation rate is similar in all three components, as is
the presence of a young stellar population. However, the position of
LSQ14mo exhibits a lower metallicity, with $12 + \log({\rm O/H}) = 8.2$ in
both the $R_{23}$ and N2 scales (corresponding to $\sim$ 0.3\,\zsun). We propose that the three bright regions in the host system are
interacting, which thus triggers star formation and forms young stellar
populations.}
 
\keywords{
supernovae: general -- supernovae: individual: (LSQ14mo),
galaxies: abundances, galaxies: dwarf, galaxies: interactions
}

\titlerunning{The interacting host galaxy of the SLSN LSQ14mo}
\maketitle

\section{Introduction}
\label{sec:introduction}

Superluminous supernovae (SLSNe) are 10-100 times brighter than normal core-collapse SNe (CCSNe) and can 
 reach absolute magnitudes of $\sim -21$ (see \citealt{2012Sci...337..927G} for a review). 
They are divided into two distinct groups from the optical spectral features around the peak brightness. The first is that of SLSNe, which do not generally exhibit hydrogen or helium lines in spectra and show no spectral signatures of interaction between fast moving ejecta and circumstellar shells, and thus are classified as SLSNe type I or type Ic \citep[e.g.][]{2010ApJ...724L..16P, 2011Natur.474..487Q, 2011ApJ...743..114C, 2013ApJ...770..128I, 2013ApJ...779...98H, 2013Natur.502..346N}. 
The lightcurves of SLSNe type I span a wide range of rise ($\sim15$-40\,d) and decline timescales ($\sim30$-100\,d). These may form two separate subclasses of slowly- and fast-declining objects (with a paucity of events at the midpoints of these ranges), but, with the current small sample size, they are also consistent with a continuous distribution \citep[see][]{2015MNRAS.452.3869N}. 
The second group is that of hydrogen-rich SLSNe (SLSNe II). A small number show broad H$\alpha$ features during the photospheric phase and do not exhibit any clear sign of interaction in their early spectra (see \citealt{2016arXiv160401226I}). These intrinsically differ from the strongly interacting SNe such as SN~2006gy \citep[e.g.][]{2007ApJ...666.1116S, 2007ApJ...659L..13O}, which are also referred to as SLSNe IIn. 
Hereafter in this paper, we only address the SLSN I types. 

Three main competing theoretical models have been proposed to explain the extreme luminosity of SLSNe of type I or Ic. These are (i) a central engine scenario, such as millisecond magnetar spin-down \citep[e.g.][]{2010ApJ...719L.204W,2010ApJ...717..245K,2012MNRAS.426L..76D} or black hole accretion \citep[e.g.][]{2013ApJ...772...30D}; (ii) pair-instability SNe (PISNe) \citep[e.g.][]{2002ApJ...567..532H} for the broad lightcurve SLSNe; and (iii) the interaction of SN ejecta with dense and massive circumstellar medium (CSM) shells \citep[e.g.][]{2011ApJ...729L...6C,2012ApJ...746..121C,2016ApJ...829...17S}.

Lightcurve modelling is an important diagnostic for current SLSNe studies. 
Firstly, the millisecond magnetar model can reproduce a wide range of lightcurves, fitting both fast and slow decliners \citep[e.g.][]{2013ApJ...770..128I, 2013Natur.502..346N}; \citet{2015MNRAS.452.1567C} found that the slowly-fading late-time lightcurve also fits a magnetar model well if the escape of high-energy gamma rays is taken into account (for time-varying leakage see \citealt{2015ApJ...799..107W}). 
In contrast, the PISN model can only explain slowly-fading lightcurve events, such as SN~2007bi, which was initially suggested to be a PISN \citep{2009Natur.462..624G} based on a $^{56}$Co decay-like tail. However, the relatively rapid rise time of similar SLSNe PTF12dam and PS1-11ap \citep{2013Natur.502..346N,2014MNRAS.437..656M} argues against this interpretation. Recently, \citet{2016MNRAS.tmp.1544K} demonstrated that radiative-transfer simulations have essential scatter depending on the input ingredients, a relatively fast rising time was found for the helium PISN model \citep{2011ApJ...734..102K}. 
Finally, the CSM model is very flexible due to having many free parameters \citep[e.g.][]{2014MNRAS.444.2096N}. However, the mechanism that could generate the dense and massive  CSM ($>$few M$_\odot$) inferred from lightcurve fitting is still unclear. The hydrogen detected in the late-time spectra of iPTF13ehe was interpreted  as CSM interaction features \citep{2015ApJ...814..108Y} but \citet{2015A&A...584L...5M} showed that interaction with a binary component was equally plausible. 

From the spectral point of view, the ionic line identifications and investigation of the excitation processes have been recently provided by spectral modelling of \citet{2016MNRAS.458.3455M} of early, photospheric 
phase spectra. The late-time nebular spectra allows us to investigate the composition of the SN ejecta. \citet{2013MNRAS.428.3227D, 2016MNRAS.455.3207J, 2017ApJ...835...13J} 
simulated the nebular spectra expected in PISN models, which show little emission in the blue region ($< 6000$\AA) of the optical spectrum because this region is fully blocked by the optically thick ejecta. 
In the red, strong \Caii $\lambda7300$ and [Fe\,{\sc i}] lines are expected along with 
neutral silicon and sulphur in the near infra-red. These model nebular spectra look markedly different from that of several slow-fading SLSNe of \cite{2017ApJ...835...13J} providing no support for a PISN interpretation. 
Instead, the similarity of the late-time nebular spectra of a number of energetic SNe Ic (e.g. SN~1998bw) and those of SLSNe has been identified by \citet{2017ApJ...835...13J, 2016ApJ...828L..18N}. 
\cite{2017ApJ...835...13J} find that at least 10\msun of oxygen is required in the ejecta to produce the 
strong [O\,{\sc i}] lines and that the material must be highly clumped. 
Moreover, \citet{2016ApJ...831...79I} presented the first spectropolarimetric observations of a SLSN, showing that SN~2015bn has an axisymmetric geometry, which is similar to those SNe that are connected with long-duration gamma-ray bursts \citep[LGRBs; e.g.][]{1998Natur.395..670G}.

The study of the host galaxies of SLSNe provide a strong constraint for understanding the stellar progenitors of SLSNe given that the distances of SLSNe ($0.1 < z < 4$; \citealt{2012Natur.491..228C}) are too far away to detect their progenitors directly.
The host galaxies of SLSNe are generally faint dwarf galaxies \citep{2011ApJ...727...15N} which tend to have low metallicity and low mass \citep{2011ApJ...730...34S, 2013ApJ...763L..28C, 2013ApJ...771...97L}.   
They share properties with LGRB host galaxies \citep{2014ApJ...787..138L,2016A&A...593A.115J}, although SLSN host galaxies appear more extreme \citep{2014ApJ...797...24V,2015MNRAS.452.1567C,2015MNRAS.449..917L,2016MNRAS.458...84A}. 
\citet{2015MNRAS.449..917L} found that the equivalent widths of \Oiii are much higher in SLSN hosts than in LGRB hosts, which they argue may imply that the stellar population is, on average, younger in SLSN than in LGRB host galaxies, and by implication that the progenitors of SLSNe are more massive than GRBs. However, this is 
not corroborated by the Hubble Space Telescope study of hosts by \citet{2015ApJ...804...90L}.
There is no significant difference in host environments of fast-declining and slowly-fading SLSNe \citep{2015MNRAS.452.1567C,2015MNRAS.449..917L,2016ApJ...830...13P}. 
Recently, \citet{2016ApJ...830...13P} and \citet{2016arXiv160504925C} both suggested that a sub-solar metallicity is required for SLSN progenitors.

The SLSN LSQ14mo was discovered by the La Silla-QUEST \citep[LSQ;][]{2013PASP..125..683B} on 2014 January 30, located at RA=10:22:41.53, Dec=$-16$:55:14.4 (J2000.0).
The Public ESO Spectroscopic Survey of Transient Objects \citep[PESSTO; ][]{2015A&A...579A..40S} took a spectrum on 2014 January 31 and identified it as a SLSN Ic \citep{2014ATel.5839....1L}. The spectrum showed a blue continuum with strong \ion{O}{ii}\, absorption features around 4200-4600 \AA, mimicking PTF09cnd at a phase of $\sim$1 week before maximum light. 
Narrow \ion{Mg}{ii}\, $\lambda \lambda$ 2796, 2803 ISM absorption lines indicated a redshift of $z = 0.253$. 
However, these lines were not resolved in the low-resolution spectrum and thus it only provided a rough redshift measurement (or the ISM is slightly in the foreground with respect to the SN location).   
In this paper, we re-identify the redshift of LSQ14mo to be $z = 0.2563$ using the narrow host emission lines (more details see Sect.\,\ref{sec:host_spectroscopy}). 
Observations of LSQ14mo were first presented by \citet{2015MNRAS.452.3869N}, who included the $r$-band and pseudo bolometric lightcurve in their statistical sample of SLSNe. Furthermore, in \citet{2015ApJ...815L..10L}, as part of the first polarimetric study of a SLSN, they found there is no evidence for significant deviations from spherical symmetry of LSQ14mo.

This paper is organised as follows: 
in Sect.\,\ref{sec:observational_data}, we give details on the photometric follow-up and spectroscopic observations of LSQ14mo and its host galaxy. We construct the bolometric lightcurve of LSQ14mo and fit models in Sect.\,\ref{sec:sn_results}, and here we also compare it to the lightcurves and spectral evolution of the other SLSNe. 
In Sect. 4, we apply the new spectral synthesis model from \citet{2016MNRAS.458.3455M} for LSQ14mo.
In Sect. 5, we present the various host galaxy properties.
In the discussion in Sects. 6 and 7, we search for other SLSNe that have spectral features indicative of a thin shell interaction; we argue that the entire host system is an interacting dwarf galaxy. Finally, we conclude in Sect. 8. The appendix contains log and magnitude tables and provides more details on data reduction. 
In this paper, we use a cosmology of $H_0 = 70$, $\Omega_{\lambda}=0.73$, $\Omega_{M}=0.27$. All magnitudes reported are in the AB system. We assumed a \citet{2003PASP..115..763C} initial mass function (IMF) of the host galaxy.

\section{Observational data}
\label{sec:observational_data}

\subsection{Supernova data}
\label{sec:lsq14mo_data}

The multi-band optical lightcurves of LSQ14mo were obtained using several telescope and instrumental configurations listed in Tables\,\ref{tab:sn_phot_opi}. These are the QUEST instrument on the ESO 1.0-m Schmidt telescope dedicated to the LSQ survey; the 2.0-m Liverpool Telescope (LT; \citealt{2004SPIE.5489..679S}) using the IO:O imager; the 6.5-m Magellan telescope and the Inamori-Magellan Areal Camera \& Spectrograph (IMACS);
the 3.58-m New Technology Telescope (NTT) using the ESO Faint Object Spectrograph and Camera (EFOSC2) in the framework of the PESSTO program \citep{2015A&A...579A..40S}
We discuss the reduction and photometric calibration of these images in detail in the Appendix.

Additionally, we have ultraviolet data from the Ultraviolet and Optical Telescope (UVOT) on the \textit{Swift} satellite, with \cite[$uvw2$, $uvm2$, $uvw1$, and $u$ observations; see][]{2008MNRAS.383..627P} covering the spectral range $\sim$ 1600-3800 \AA. Those observations were acquired only within a few days of maximum light.
The data were reduced using the HEASARC \footnote{NASA's High Energy Astrophysics Science Archive Research Center} software.  All measured magnitudes and  detection limits are listed in Table\,\ref{tab:sn_phot_opi} and \ref{tab:sn_phot_uv}.

We obtained a series of spectra of LSQ14mo from $-7.0$d to +56.5d using NTT+EFOSC2 within the PESSTO program. 
The observational log is reported in Table\,\ref{tab:sn_spec}. 
The NTT + EFOSC2 data were reduced in a standard fashion using the PESSTO pipeline \citep{2015A&A...579A..40S}. 
This applies bias-subtraction and flat-fielding with halogen lamp frames. 
The spectra were extracted using the pipeline and wavelength-calibrated by identifying lines of He and Ar lamps, and flux-calibrated using sensitivity curves obtained using spectroscopic standard star observations on the same nights. 
The final data products can be found in the ESO Science Archive Facility as part of PESSTO SSDR2\footnote{Data access described on www.pessto.org}, and all spectra will also be available through WISeREP\footnote{http://wiserep.weizmann.ac.il} \citep{2012PASP..124..668Y}, and along with photometry through the Open SN Catalog\footnote{https://sne.space} \citep{2017ApJ...835...64G}.

\begin{table*}
 \begin{minipage}{190mm}
 \centering
  \caption{Log of spectroscopic observations of LSQ14mo and its host galaxy.  The phase (day) has been corrected for time dilation ($z = 0.256$) and relative to the SN {\it r}-band maximum on MJD 56697. Resolution of NTT + EFOSC2 is adopted from \citet{2015A&A...579A..40S}, as measured from the sky lines. For VLT + FORS2 with a filter {\it GG435+81} and a 1\arcsec.0 slit, the resolving power is 440 at the central wavelength 5900\AA, and the dispersion is 3.27 \AA/pixel. }
\label{tab:sn_spec}
\begin{tabular}[t]{llllllllll}
\hline\hline 
Date & MJD & Phase & Telescope & Instrument & Grism & Exp. time   & Slit  & Resolution  & Range   \\
  &   &  (day) &   &   &  &   (sec) &  (\arcsec) &  (\AA) &   (\AA) \\
\hline
2014 Jan 31     & 56688.16      & -7.0          & NTT & EFOSC2 & Gr\#13 & $1800\times1$   & 1.0 & 18.2  &  3668-9269\\
                        &                       &               &          &                 & Gr\#11 & $1800\times2$  & 1.0 &  13.8 &  3357-7486\\
2014 Feb 6      & 56694.19      & -2.2          & NTT & EFOSC2 & Gr\#11 & $1800\times3$   & 1.0 & 13.8 &  3357-7486\\
        &       &       &   &   & Gr\#16 & $1800\times3$        & 1.0 & 13.4 &  6008-10008\\
2014 Feb 20     & 56708.06      & 8.8   & NTT & EFOSC2 & Gr\#13 & $1800\times3$         & 1.0 & 18.2 &  3668-9269\\
2014 Feb 28     & 56716.13      & 15.2  & NTT & EFOSC2 & Gr\#13 & $1800\times3$         & 1.0 & 18.2 &  3668-9269\\
2014 Mar 8      & 56724.26      & 21.7  & NTT & EFOSC2 & Gr\#13 & $2400\times3$         & 1.0 & 18.2 &  3668-9269\\
2014 Apr 21     & 56768.04      & 56.5  & NTT & EFOSC2 & Gr\#13 & $2400\times4$         & 1.0 & 18.2 &  3668-9269\\
2015 Jan 23     & 57045.18      & 277.2 & VLT  & FORS2 & GRIS$\_$300V & $1230\times10$ & 1.0 & 10.2 & 4300-9600\\
2015 Feb 21     & 57074.28      & 300.4 & VLT  & FORS2 & GRIS$\_$300V & $1230\times3$ & 1.0 & 10.2 & 4300-9600\\
2015 Feb 24     & 57077.27      & 302.8  & VLT  & FORS2 & GRIS$\_$300V & $1230\times2$ & 1.0 & 10.2 & 4300-9600\\
\hline 
\end{tabular}
\end{minipage}
\end{table*}

\subsection{Host galaxy photometry}
\label{sec:host_photometry}

\begin{figure}
\centering
\includegraphics[angle=0,width=\linewidth]{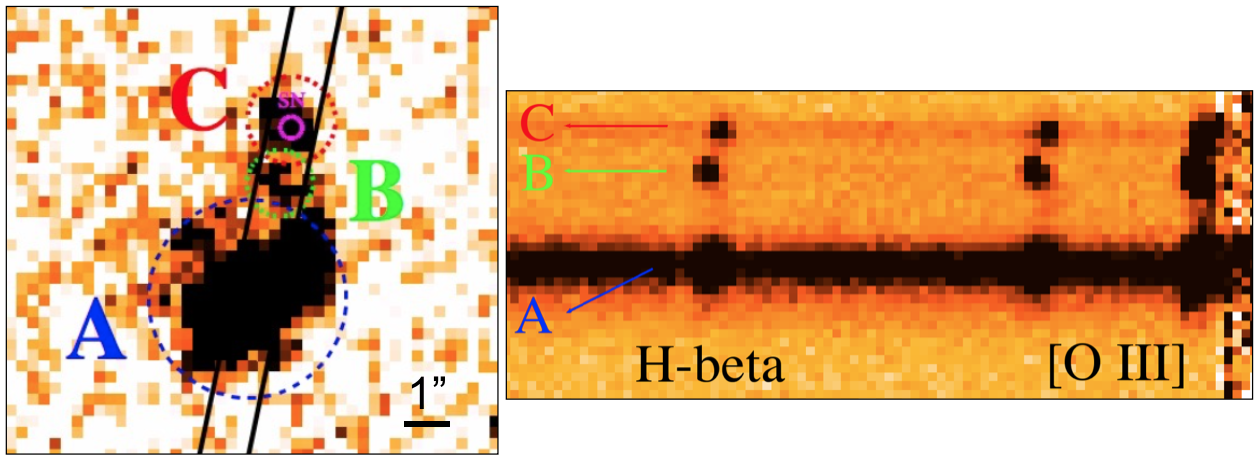}
\caption{{\it Left:} VLT + FORS2 slit angle overlaid on NTT + EFOSC2 (PESSTO) $i$-band image of the host galaxy system taken on 2015 February 12 (+293.1\,d). We have labelled the LSQ14mo position (C), bright \Hii region (B) and PSO J155.6730-16.9216 (A). North is up and east is left. {\it Right:} Portion of the 2D spectrum from VLT + FORS2 showing the emission lines \hb and \Oiii$\lambda\lambda$4959,5007. Three spatially resolved components, A, B, and C, also offset in velocity, are clearly seen.}
\label{fig:vlt_slit_position}
\end{figure}

We gathered deep, high-spatial-resolution images of LSQ14mo and its host galaxy system after +250\,d with PESSTO (see Sect.\,\ref{sec:SN_photometry}).
We chose the best-seeing (0.8\arcsec) $i$-band image taken at +293.1\,d to determine that there are three components in the host galaxy system PSO J155.6730-16.9216. We have named these positions A, B, and C from south to north, respectively (see Fig.\,\ref{fig:vlt_slit_position}) and use these names throughout this paper. The SN is located coincident with the flux at position C. 
We selected appropriate aperture sizes for positions A, B, and C to limit the contamination from each corresponding component (2.5\arcsec, 0.75\arcsec\ , and 1\arcsec\ for positions A, B, and C, respectively). 
Aperture photometry was then carried out within {\sc iraf/daophot}, and we used the same aperture size to measure the flux of local secondary standards in the field (see Sect.\,\ref{sec:SN_photometry}) to set the zeropoint. We found $i = 20.98\pm 0.04$\,mag for component A and $23.75\pm 0.10$\,mag and $23.44\pm 0.10$\,mag for components B and C, respectively.
We obtained pure host images on 2015 December 17 (+539.2\,d) using the Gamma-Ray Burst Optical/Near-Infrared Detector \citep[GROND;][]{2008PASP..120..405G}. We did not detect a 4000\AA\ break for positions B and C, expected to have $g - r > 0.3$ mag from the synthetic photometry of the spectra.    
All photometry measurements are listed in Table\,\ref{tab:host_phot_grond}.

The brightest component, position A, was detected in Pan-STARRS1 $3\pi$ images 
as PSO J155.6730$-$16.9216. Those stack images (from February 2010 to 2013 May) were taken before the SN explosion.
The position A was also marginally detected in Galaxy Evolution Explorer (GALEX) images and designated as GALEX J102241.6$-$165517, with a magnitude of NUV = $22.87 \pm 0.39$ ($\sim$ 2271 \AA) in the AB system (GR6 catalogue\footnote{http://galex.stsci.edu/GR6/}).

\begin{table*}
\centering
\caption{Photometry of each component of the host galaxy system of LSQ14mo using GROND, data were taken on 2015 December 17 (MJD = 57374.28), at +539.2\,d. All magnitudes are in the AB system and the errors include statistical and systematic errors.
The aperture reported was measured from the $r'$-band image.}
\label{tab:host_phot_grond}
\begin{tabular}{|l|l|l|l|l|l|l|l|}
\hline\hline 
Position (aperture) & $g$ & $r$ & $i$ & $z$ & $J$ & $H$ & $K$ \\
\hline
A (1.9\arcsec) & 21.85   (0.15) & 21.27 (0.02) & 21.11 (0.05) & 21.07 (0.04) & 21.18 (0.09) & 21.31 (0.17) & 20.65 (0.14) \\
B (0.6\arcsec) & 23.67 (0.15) & 23.59 (0.06) & 23.59 (0.12) & 23.67 (0.18) & - & - & - \\          
C (0.8\arcsec) & 23.82 (0.15) & 23.61 (0.06) & 23.50 (0.12) & 23.56 (0.18) & 23.61 (0.62) & - & - \\               
A + B (3.2\arcsec) & 21.49 (0.16) & 21.00 (0.03) & 20.97 (0.05) & 20.92 (0.06) & 20.79 (0.09) & - & - \\       
\hline
\end{tabular}
\end{table*}

\begin{figure}
    \centering
    \begin{subfigure}[h]{0.85\columnwidth}
        \includegraphics[width=\columnwidth,trim={1.5cm 0cm 1.5cm 0.25cm}]{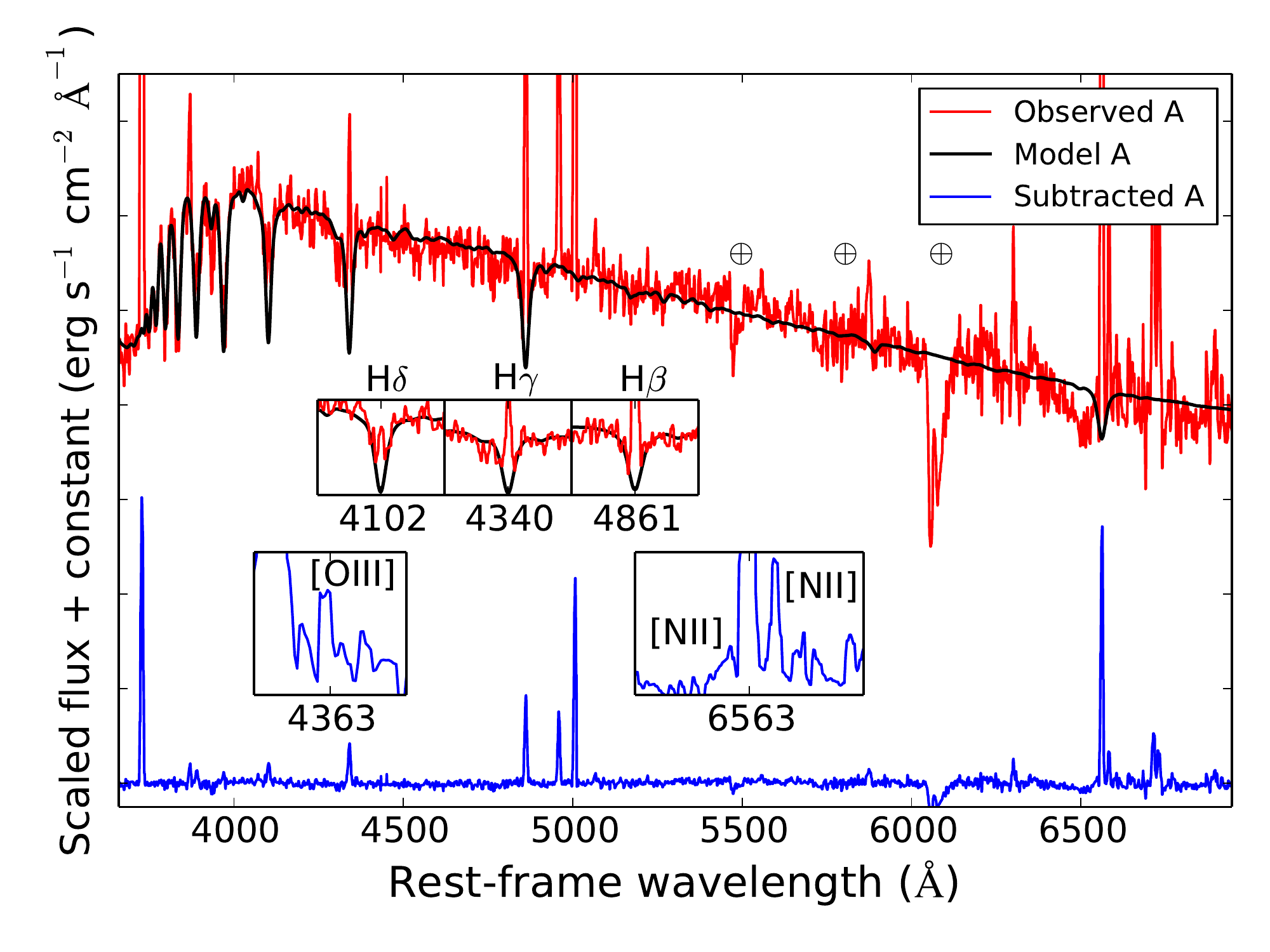}
        \caption{Position A.}
        \label{fig:first_sub}
    \end{subfigure}  
    \begin{subfigure}[h]{0.85\columnwidth}
        \includegraphics[width=\columnwidth,trim={1.5cm 0cm 1.5cm 0.25cm}]{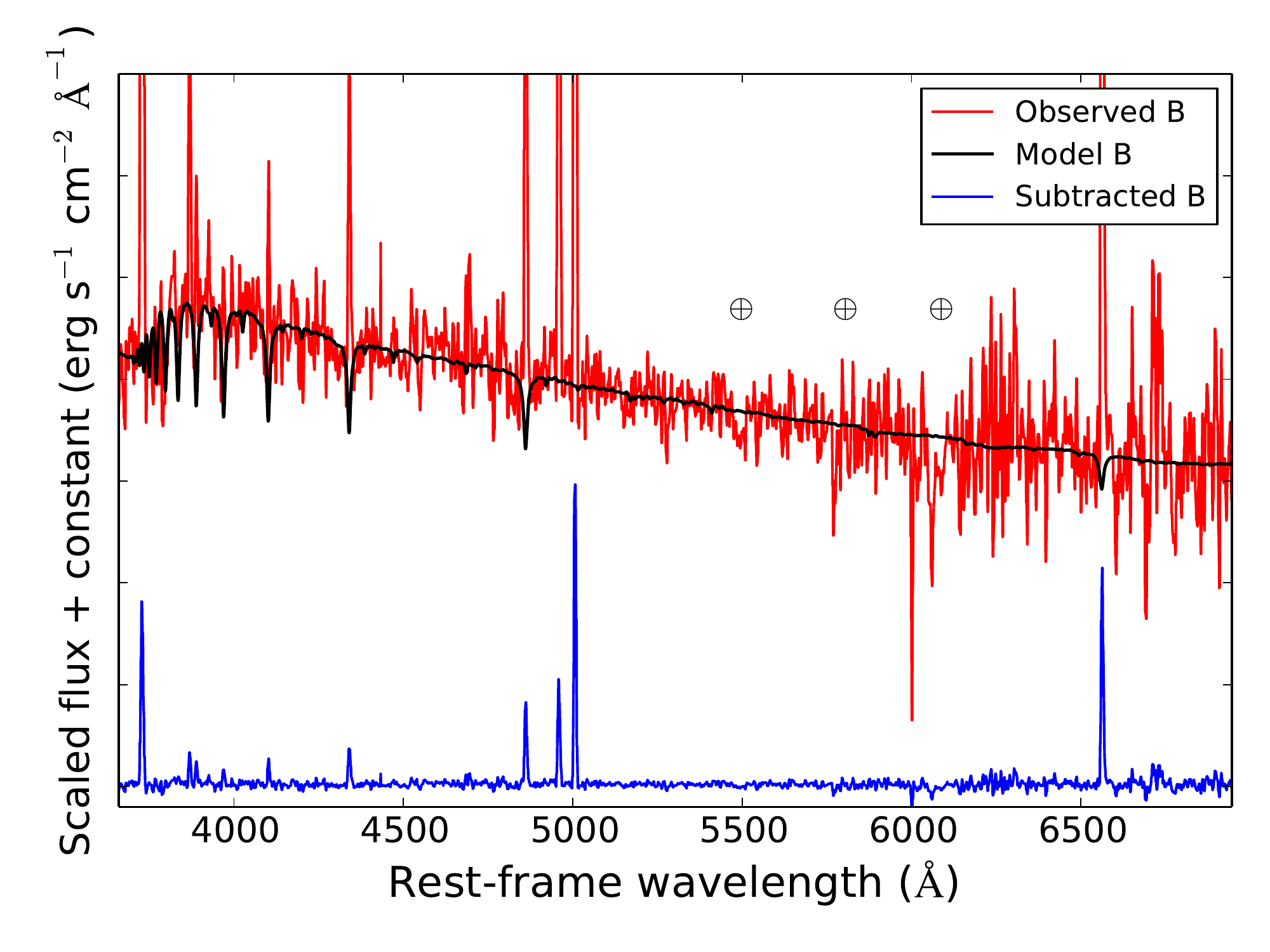}
        \caption{Position B.}
        \label{fig:second_sub}
    \end{subfigure}
    \begin{subfigure}[h]{0.85\columnwidth}
        \includegraphics[width=\columnwidth,trim={1.5cm 0cm 1.5cm 0.25cm}]{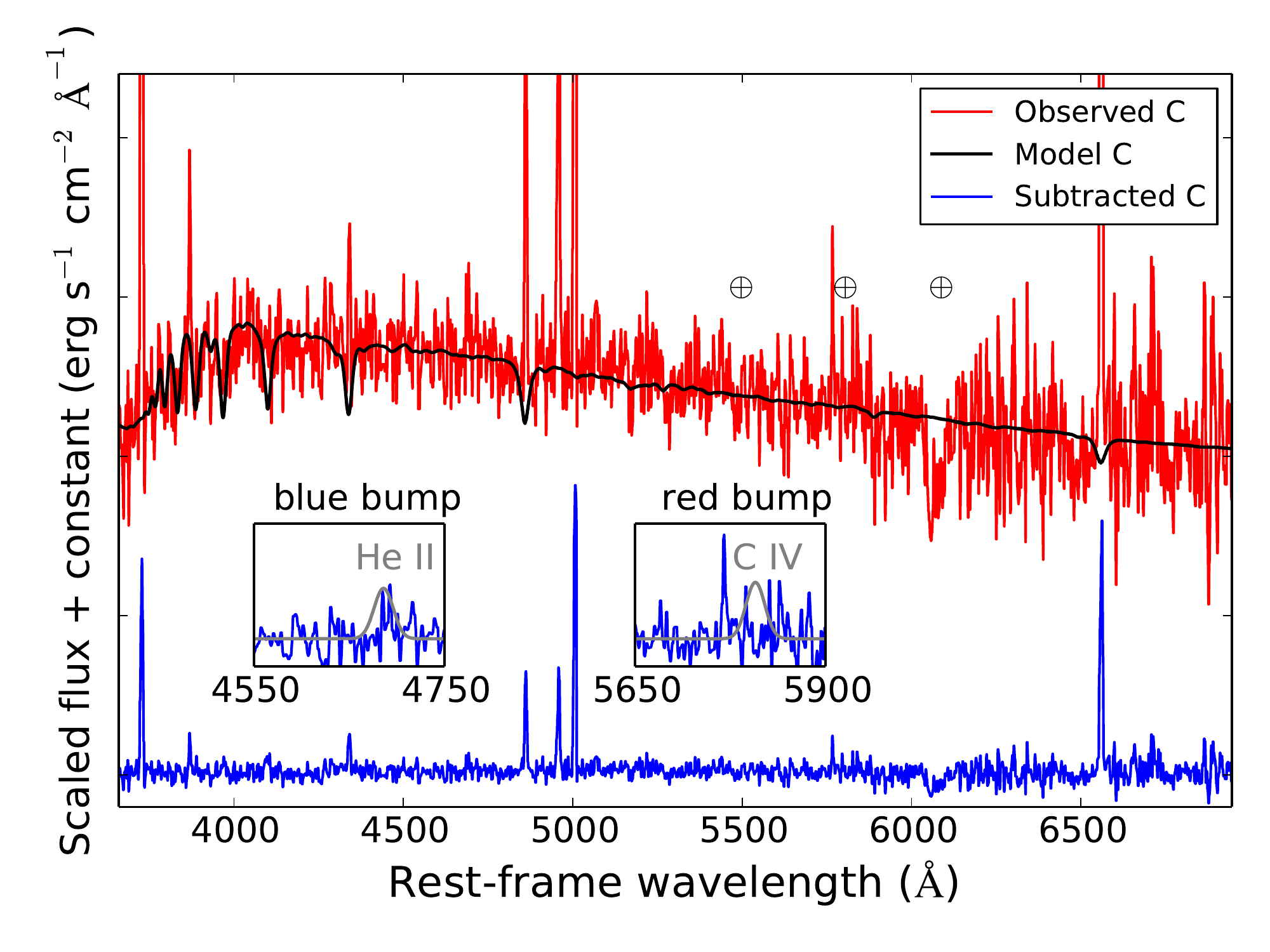}
        \caption{Position C.}
        \label{fig:third_sub}
    \end{subfigure}  
    \caption{VLT + FORS2 spectra of the host galaxy system extracted at the three spatially resolved positions (A, B, and C; shown in red) and fit with the stellar population models from {\sc starlight} (black lines overlapping observed spectra; see Sect.~\ref{sec:line_measurements}). The blue spectra are subtracted residuals for the models. For position A, the upper inset panels show strong stellar absorption features of Balmer lines, while the bottom inset panels show detections of a weak auroral $\lambda4363$ line and clear detections of \Nii lines around \ha. For position C, the inset panels zoom in to WR feature regions, and no broad feature was detected.
The atmospheric telluric absorption features are marked with the earth symbol.}
\label{fig:host_model_spec}
\end{figure}

\subsection{Host galaxy spectroscopy}
\label{sec:host_spectroscopy}

The deep host spectra were taken with the 8.2m Very Large Telescope (VLT) + FOcal Reducer and low dispersion Spectrograph (FORS2; \citealt{1998Msngr..94....1A}) during 2015 January and February (approximately +294\,d, listed in Table\,\ref{tab:sn_spec}). The slit angle is 13.066 degrees rotated from the north (see Fig.\,\ref{fig:vlt_slit_position} for the slit position) to cover the C (SN location) and A positions. We shifted the slit position by $\sim$ 40 pixels for each individual frame, allowing us to remove the sky emission lines by two-dimensional (2D) image subtraction. 

The 2D images were reduced using standard techniques within {\sc iraf} for bias subtraction and then removed cosmic rays using {\sc l.a.cosmic} \citep{2001PASP..113.1420V}, an algorithm for robust cosmic ray identification and rejection. We finally stacked those 15 frames using an inverse-variance weighted average. 
The stacked image (Fig.\,\ref{fig:vlt_slit_position}) clearly shows emission lines at three resolved spatial positions. The separations between those emission lines are consistent with the brightest regions at positions A, B, and C in the deep NTT + EFOSC2 $i$-band image. 

We extracted 1D spectra from the stacked science frame using the {\sc iraf/longslit} package.  
We took a spectrophotometric standard LTT 2415 for the flux calibration and used daytime He+HgCd+Ar arcs for the wavelength calibration. More details are given in the Appendix.
Subsequently, we corrected observed spectra for a Milky Way extinction of $A_{V} = 0.20$\,mag, which, for the assumed $R_{V}$ = 3.1, corresponds to $E(B-V) = 0.06$\,mag \citep{2011ApJ...737..103S}. We then shifted the spectra to the rest-frame: $z = 0.2556$ for spectrum A, $z = 0.2553$ for spectrum B and $z = 0.2563$ for spectrum C. The redshift was estimated based on the observed wavelength of \hb and \Oiii $\lambda\lambda$4959, 5007 lines. The slightly different redshifts of three components are clearly indicated from the emission line positions, see Fig.\,\ref{fig:vlt_slit_position}. Fig.\,\ref{fig:host_model_spec} shows the final spectra corresponding to positions A, B, and C.

We estimated radial velocity offsets from their relative redshifts:  
$\sim -90$\,km\,s$^{-1}$ from position A to B, $\sim 300$\,km\,s$^{-1}$ from B to C, and $\sim 210$\,km\,s$^{-1}$ from A to C. 
For comparison, the rotational velocity of the Milky Way is approximately 220 km\,s$^{-1}$ at the position of the Sun. The radial velocities are approximately 270 km\,s$^{-1}$ for the Large Magellanic Cloud (LMC) and 150 km\,s$^{-1}$ for the Small Magellanic Cloud (SMC) \citep[see Fig. 3 of][]{2005A&A...432...45B}. 
Hence, these three components may be kinematically associated, although it is not possible to distinguish whether the three components are part of a single galaxy with B and C being tidal tails of A, or if C is a satellite galaxy of A. 
Therefore, we refer to them as a ``host galaxy system'' in this paper. 
For further discussion of the nature of this host system, we refer to Sect.\,\ref{sec:discussion_host}, where we propose that the component B is a new-born \Hii region and the component C may be a satellite or merging with
 the component A.

\section{Supernova results}
\label{sec:sn_results}

\subsection{Lightcurve}
\label{sec:sn_results_lc}

\begin{figure}
\includegraphics[angle=0,width=\linewidth]{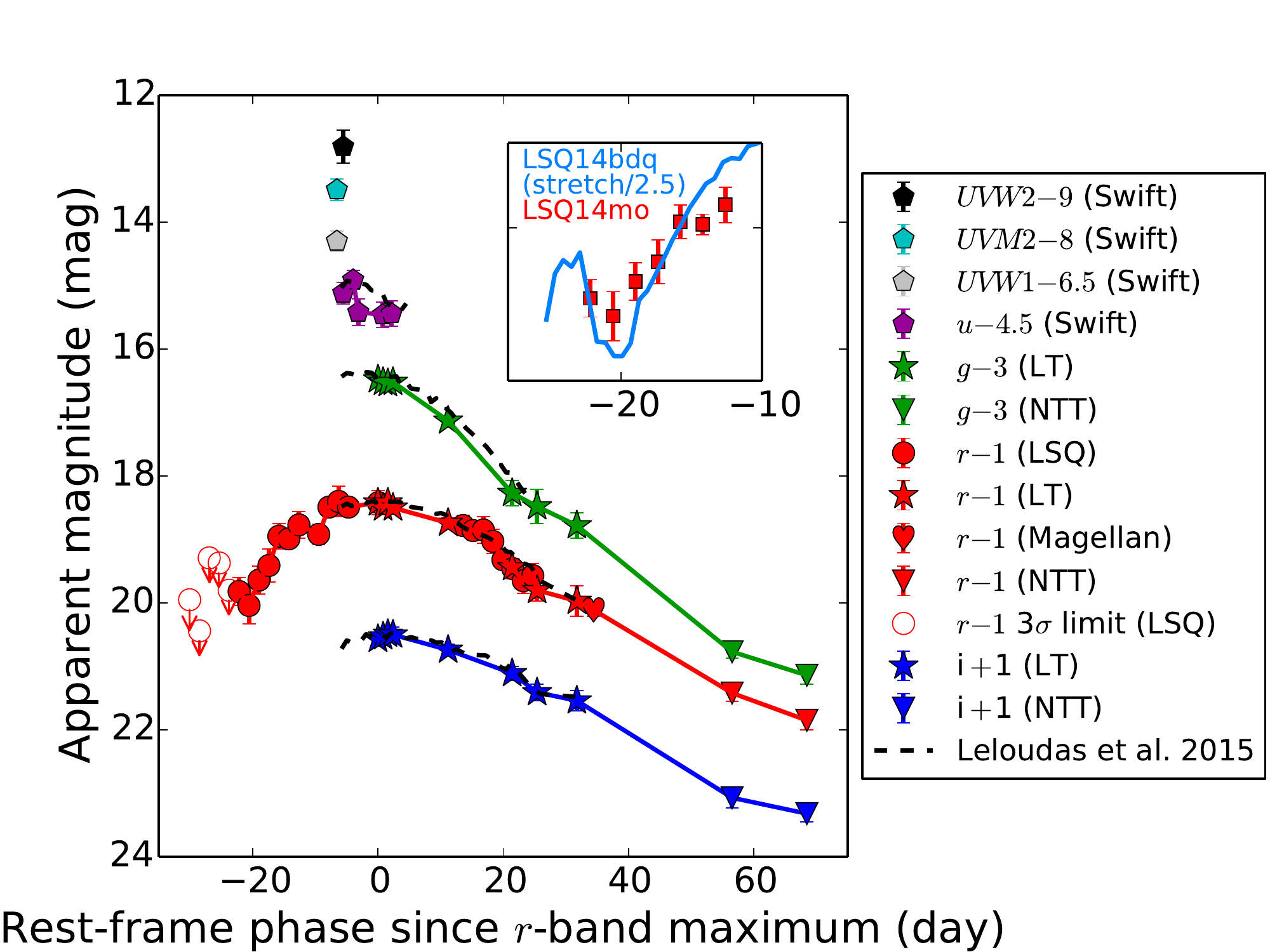}
\caption{Photometry of LSQ14mo from the UV to the optical. The early- and late-time $3\sigma$ detection limits are shown in empty symbols. 
The phase (day) has been corrected for time dilation ($z = 0.256$) and relative to the SN {\it r}-band maximum on MJD 56697. The period from 100 to 200 days is skipped due to no observations. The inset panel shows the absolute $g$-band pre-maximum lightcurves of LSQ14mo and LSQ14bdq with a stretch in time divided by 2.5, an early bump of LSQ14mo is plausibly detected.
For comparison, the optical photometry given by \citet{2015ApJ...815L..10L}, which is plotted as dashed lines, is in a good agreement with our measurements.}
\label{fig:sn_phot_obs}
\end{figure}

\begin{figure}
\includegraphics[angle=0,width=\linewidth,trim={0.5cm 0cm 1cm 1cm}]{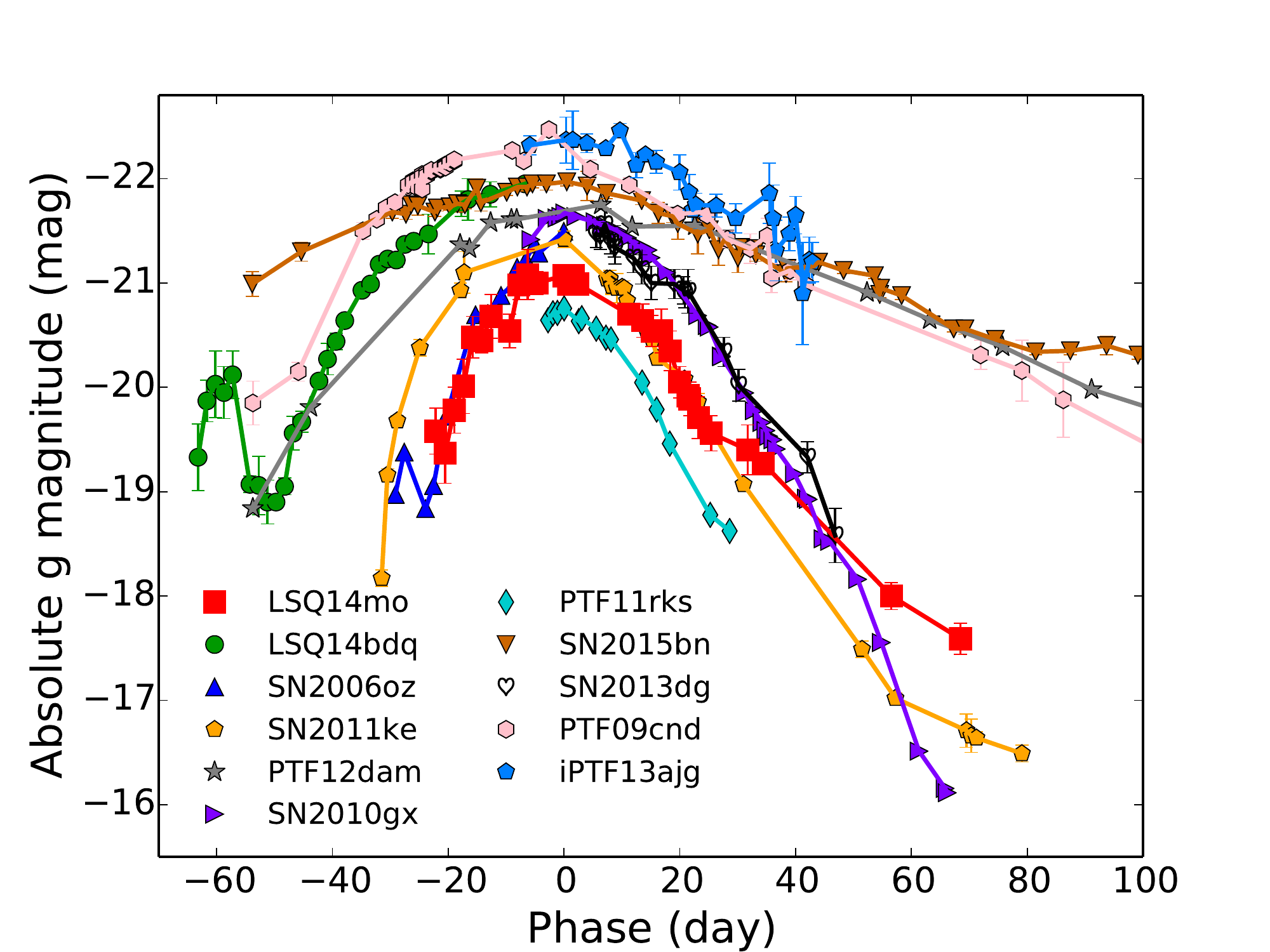}
\caption{Absolute rest-frame $g$-band lightcurve of LSQ14mo compared with other type I SLSNe. The peak brightness of LSQ14mo of $-21$ mag is typical for fast-declining events and fainter than slowly-fading SLSNe. We note that after $+50$\,d, LSQ14mo and SN~2011ke show a shallower decline rate, which is similar to a $^{56}$Co decay tail. 
Data for comparison are taken from \citet{2010ApJ...724L..16P,2011Natur.474..487Q,2012A&A...541A.129L,2013ApJ...770..128I,2013Natur.502..346N,2015ApJ...807L..18N,2016ApJ...826...39N,2014ApJ...797...24V}.}
\label{fig:sn_phot_rf}
\end{figure}

Fig.\,\ref{fig:sn_phot_obs} shows the observed lightcurves of LSQ14mo. We have ground-based photometry of LSQ14mo in $gri$ bands covering the period $-20$\,d to +70\,d, and late-time ($>250$\,d) detection limits. We also have a few epochs around maximum light from the 
\textit{Swift} UV observations. The comparison of our data with the early-phase photometry of \citet{2015ApJ...815L..10L} shows excellent agreement.
The two-day cadence survey strategy employed by LSQ provides a good constraint on the explosion epoch. The earliest detections of LSQ14mo appear to show an initial decline, which could be consistent with the double-peaked structure exhibited by LSQ14bdq \citep{2015ApJ...807L..18N}, SN~2006oz \citep{2012A&A...541A.129L}, DES14X3taz \citep{2016ApJ...818L...8S} and PTF12dam \citep{2017ApJ...835...58V}, and has been proposed to be common in SLSNe Ic \citep{2016MNRAS.457L..79N}. However, the photometric errors on these points are relatively large, and an early bump cannot be definitively identified in the case of LSQ14mo. The inset panel of Fig.\,\ref{fig:sn_phot_obs} shows the absolute $g$-band pre-maximum lightcurves of LSQ14mo and LSQ14bdq with a stretch in time divided by 2.5.

Our best-sampled photometry is in the $r$ band, which, at redshift $z = 0.256,$ approximately corresponds to the rest-frame $g$ band. We converted to absolute magnitudes using a distance modulus from the cosmology calculator of \citet{2006PASP..118.1711W}, corrected for a Milky Way extinction ({\it A$_{V}$} = 0.201), and applied {\it K}-corrections. 
We employed the {\sc pysynphot} module within {\sc python} to calculate synthetic photometric quantities for spectra at the given $gri$-band pass, then we shifted the spectra to the rest-frame ($z = 0.256$) and calculated the synthetic photometry again. The differences between these were adopted as {\it K}-correction values, which were double-checked using the {\sc snake} code \citep{2016arXiv160401226I}. These values are given in Table\,\ref{tab:sn_K-correction}. We interpolated the corrections linearly to all epochs with imaging. Finally, we considered that the internal dust extinction from the host galaxy is negligible given by the ratio of \ha and \hb lines.

In Fig.~\ref{fig:sn_phot_rf}, we compare the rest-frame $g$-band lightcurve to a sample of other SLSNe. The peak luminosity, $M_g \approx -21$, is typical of these fast-declining events and fainter than those slowly-fading objects \citep[for a larger sample comparison, we refer to][]{2015MNRAS.452.3869N}. The rise time and overall lightcurve width are very similar to SN~2011ke \citep{2013ApJ...770..128I}, SN~2006oz \citep{2012A&A...541A.129L}, and SN~2010gx \citep{2010ApJ...724L..16P} and much narrower than other events, such as PTF12dam \citep{2013Natur.502..346N,2015MNRAS.452.1567C} and LSQ14bdq \citep{2015ApJ...807L..18N}. LSQ14mo appears to transition to a shallower decline rate after $\approx30$\,d from peak. Similar transitions are apparent in other SLSNe, notably SN~2011ke, but the tail of LSQ14mo is brighter than SN~2011ke, and by +70\,d it is much brighter than SN~2010gx, highlighting the rather diverse late-time behaviour of SLSN lightcurves.

\subsection{Spectroscopic evolution}
\label{sec:sn_results_spec}

\begin{figure}
\includegraphics[angle=0,width=\linewidth]{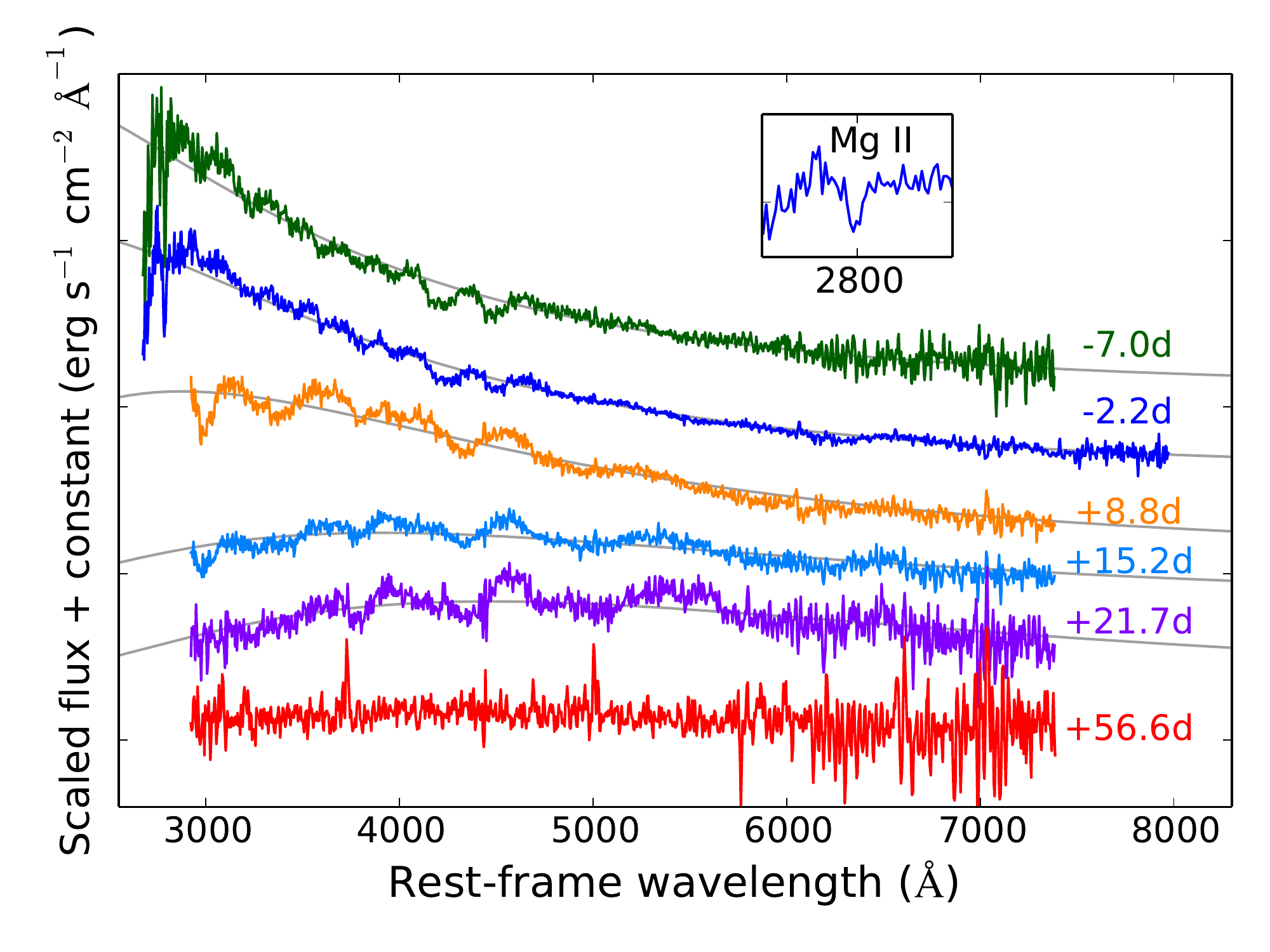}
\includegraphics[angle=0,width=\linewidth]{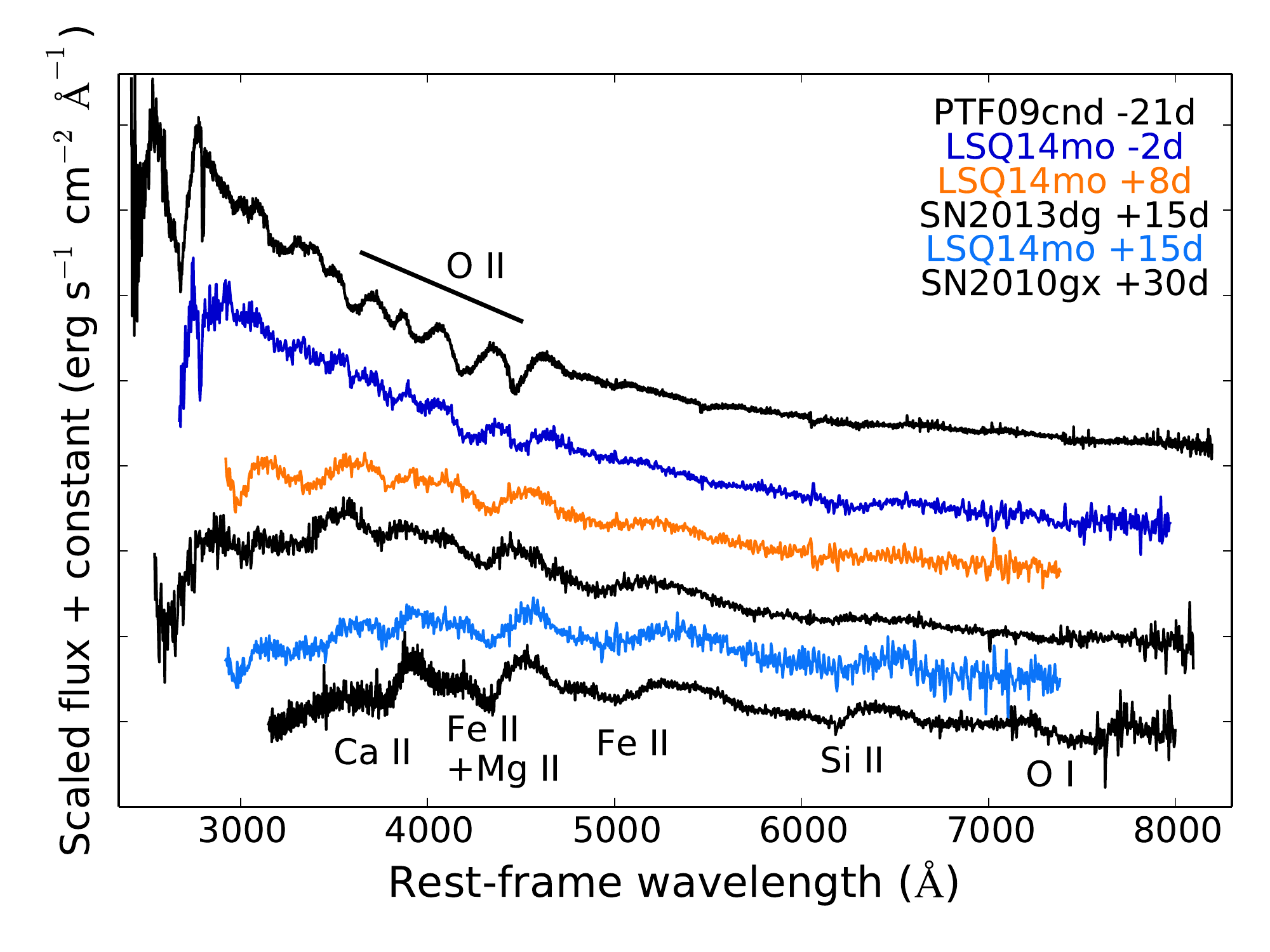}
\caption{{\it Top:} Spectral evolution of LSQ14mo taken by NTT+EFOSC2 of the PESSTO program. Blackbody curve fits (grey) show that the temperature before maximum light is $\approx 13000$-15000\,K and cooling rapidly to $\sim$7000\,K by 20 days after peak. The inset shows the \Mgii absorption features from the host ISM taken at $-2.2$d.
{\it Bottom:} Comparison with other SLSN spectra. The prominent lines and their profiles are virtually identical to other SLSNe.  } 
\label{fig:sn_spec}
\end{figure}

The NTT+EFOSC2 (of the PESSTO program) spectra of LSQ14mo are shown in Fig.~\ref{fig:sn_spec}. They have been corrected for Galactic extinction and shifted to the rest-frame at $z = 0.256$. The pre-maximum spectra are dominated by a blue continuum with a blackbody colour temperature of $\approx13000$-15000\,K, and a distinctive W-shaped absorption feature between 4100 and 4500\,\AA. These lines have been attributed to O\,\textsc{ii} by many authors since they were first identified by \citet{2011Natur.474..487Q}, and are generally seen in SLSNe Ic at this phase. The presence of ionised oxygen is consistent with the high temperatures around the peak. Recently, \citet{2016MNRAS.458.3455M} suggested that these lines must be excited non-thermally. One scenario that could explain this is X-ray injection from a magnetar.

As LSQ14mo ages, its spectrum resembles various other well-observed SLSNe, such as PTF09cnd \citep{2011Natur.474..487Q}, SN 2013dg \citep{2014MNRAS.444.2096N}, and SN 2010gx \citep{2010ApJ...724L..16P}. The continuum becomes substantially cooler between one and two weeks after maximum and the black-body fitting temperature drops from $\sim 14000$\,K at peak to $\sim 7400$\,K at $+15.2$\,d.
At this phase, it is dominated by Fe\,\textsc{ii} and intermediate-mass elements (bottom panel of Fig.~\ref{fig:sn_spec}). In Sect. \ref{sec:synthesis_spec}, we examine the evolution of the lines and continuum in more detail using spectral synthesis models \citep{2016MNRAS.458.3455M}.

\subsection{Bolometric lightcurve and model fits}
\label{sec:sn_results_bol_model}

To build the bolometric lightcurve of LSQ14mo, we used our photometry and spectroscopy to
construct the spectral energy distribution (SED) as a function of time, and built a bolometric lightcurve with a code {\sc superbol}\footnote{https://github.com/mnicholl/astro-scripts/blob/master/superbol.py}. 
As we lack coverage in the NIR and have \textit{Swift} UV photometry only at maximum light, we estimated the UV and NIR contribution at each epoch with spectroscopy via blackbody fits. Since blackbody fits do not take into account metal absorption lines in the UV \citep[e.g.][]{2011ApJ...743..114C}, we used our \textit{Swift} measurements to calculate the degree of absorption. Around $-5$d, we found that the flux in the $UVW2$ (filter central wavelength at 1928\AA) is 48\% of the blackbody prediction, and 56\% in the $UVW1$ (filter central wavelength at 2600\AA). We assumed that this suppression ratio is constant at all times.

We then linearly interpolated the $K$-corrected $g$- and $i$- band -- and predicted UV and NIR -- fluxes to all epochs with $r$-band photometry, corrected for the distance to the SN, and numerically integrated the SED assuming that the flux goes to zero outside of the \textit{Swift} $uvw2$ and $K$ bands. We also added additional 4\% flux from the UV contribution for the last two epochs, because, as discussed by \citet{2015MNRAS.452.1567C}, UV radiation is not zero in the late time. 
The fraction of energy emitted in the optical, UV, and NIR for LSQ14mo is very similar to other fast-declining SLSNe \citep{2013ApJ...770..128I}. At $-20$\,d, the fraction of UV is 41\%, optical is 53\%, and NIR is 6\%; at $+20$\,d, those fractions evolved to 12\% UV, 69\% optical, and 19\% NIR; and at $+70$\,d, these fractions are 4\% UV, 60\% optical, and 36\% NIR.  
This gives us confidence in the method used to derive the fluxes from $UVW2$ through to $K$ band, but does not account for flux beyond these limits. 
For comparison, we also present the optical pseudo-bolometric lightcurve as well, based on real $ugri$ photometry. 

\begin{figure}
       \centering         
       \includegraphics[angle=0,width=\linewidth,trim={0 0 1cm 1cm}]{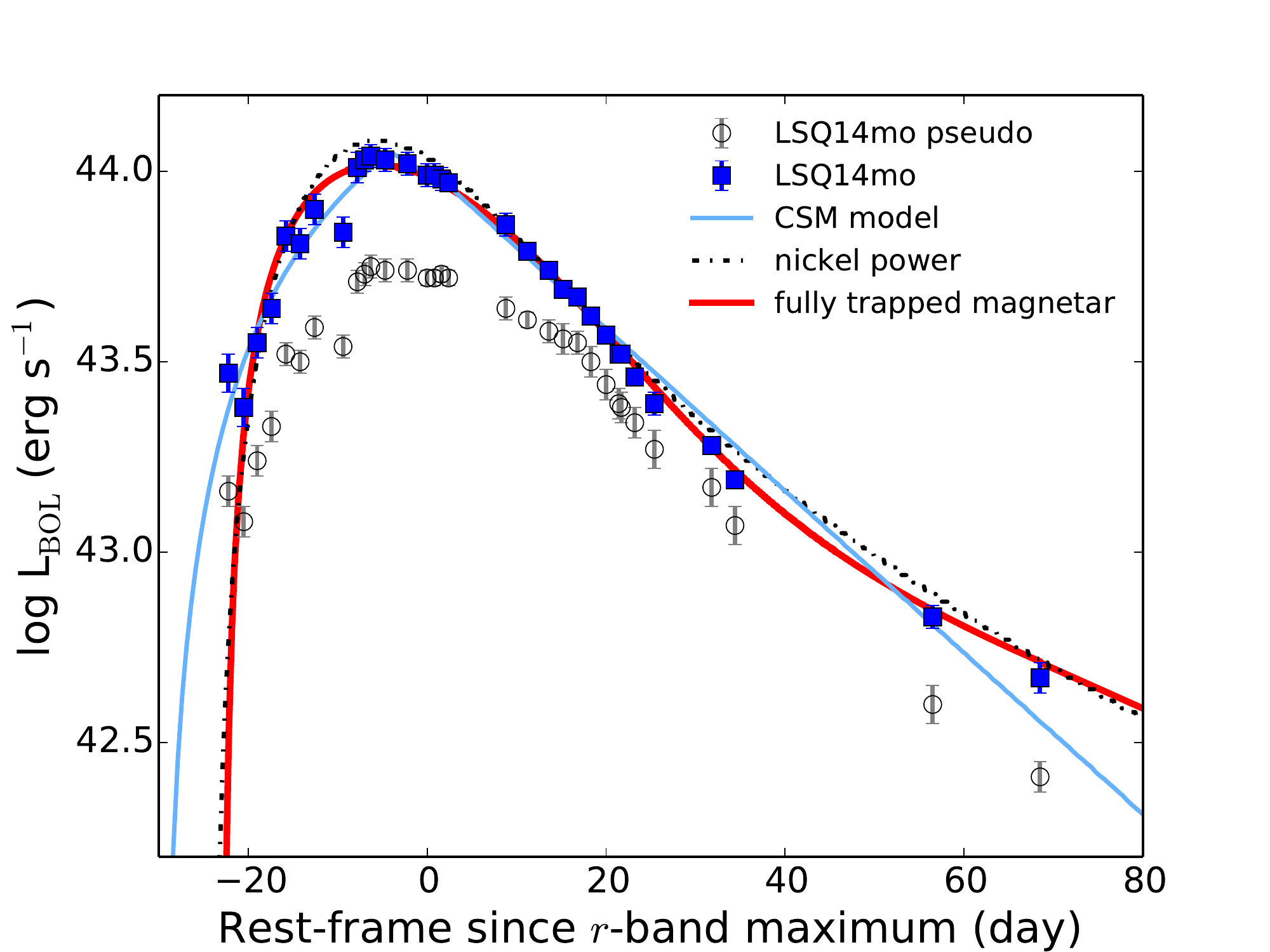}
\caption{Bolometric lightcurves of LSQ14mo and alternative model fits. Black empty circles show the pseudo bolometric lightcurve from $ugri$ photometry and blue squares show the final bolometric lightcurve integrated from $UVW2$ to $K$. The fully trapped magnetar model shown with a red curve indicates a magnetar with a magnetic field of $5.1\times 10^{14}~\mathrm{G}$ and an initial spin period of 3.9~ms and an SN ejecta mass of $M_{\rm ej} = 3.9~\msun$. The Ni-powered model requires 7~\msun of $^{56}$Ni, which is $\sim 90$\% of the SN ejecta mass.
The interacting model gives a SN ejecta of $M_{\rm ej}=3.3$\,\msun and a dense ($\rho_{\rm CSM}=7\times10^{-13}$\,g\,cm$^{-3}$) CSM of $M_{\rm CSM}=2.0$\,\msun.}
\label{fig:sn_bol_model}
\end{figure}

The pseudo bolometric lightcurves and full bolometric lightcurves of LSQ14mo are plotted in Fig.\,\ref{fig:sn_bol_model} and the data points are listed for reference in Table\,\ref{tab:bol_flux}.  
Uncertainties are accounted for in the error bars in Table\,\ref{tab:bol_flux}. These include the error in photometric measurements and an uncertainty arising from the extrapolation methods for the missing filter data, using constant colour evolution or linear interpolation. The luminosity limits derived from our late non-detections are unfortunately not deep enough to provide constraints to the model, hence we do not plot these on the figures.

\begin{figure}
       \centering         
       \includegraphics[angle=0,width=\linewidth,trim={0 0 0.5cm 0.5cm}]{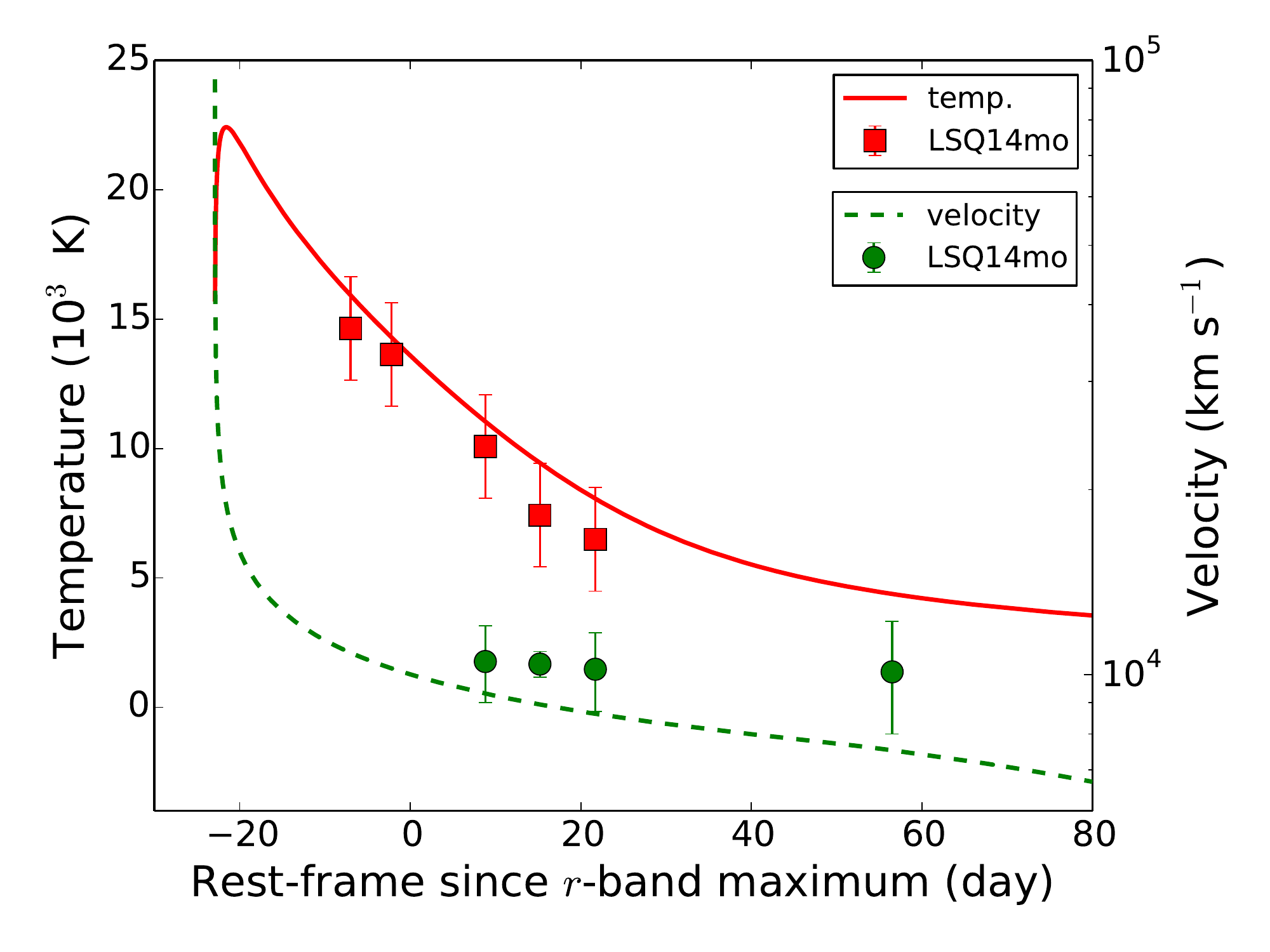}
\caption{Temperature and velocity evolution of LSQ14mo. Red and green lines are given from the
magnetar modelling results. For observational data, the velocity was measured from the \Feii $\lambda5169$ line, and temperature was measured by fitting a black-body curve to spectra. The consistency of those two independent parameters is suggestive of a magnetar as the power source.}
\label{fig:sn_magnetar_tem_v}
\end{figure}

As shown by Fig.~\ref{fig:sn_bol_model}, the magnetar spin-down models \citep{2013ApJ...770..128I}\footnote{Magnetar lightcurve fitting code is available at  A.~Jerkstrand's webpage: https://star.pst.qub.ac.uk/webdav/public/ajerkstrand/Codes/Genericarnett} provide an excellent fit to the data. We assumed that the magnetar spin-down energy is fully trapped in the ejecta and obtained a best fit (reduced $\chi^{2}$ of 1.40) with a magnetic field strength of $5.1\times 10^{14}~\mathrm{G}$ and  initial spin period of 3.9~ms with $M_{\rm ej} = 3.9~\msun$ and $E_{\rm k} = 3.14\times10^{51}~\mathrm{erg}$. 
Figure\,\ref{fig:sn_magnetar_tem_v} shows the evolution of temperature and velocity of LSQ14mo. The results 
from the magnetar model fit are compared with observational data: the velocity was measured from the
 minimum of \Feii $\lambda5169$ absorption and the temperature was measured by fitting a black-body curve to spectra.
Those independent parameters from the model and from the observation got a good agreement, which supports a magnetar as the underlying power source of LSQ14mo.

Alternatively, we can fit the data with an interaction-powered model \citep{2012ApJ...746..121C}. We assume a uniform CSM shell with an inner radius of $10^{15}$\,cm \citep[for details see][]{2014MNRAS.444.2096N}. Our best-fit model is shown in Fig.~\ref{fig:sn_bol_model}, and reproduces the data very well. The parameters for this fit are: $M_{\rm ej}=3.3$\,\msun, $M_{\rm CSM}=2.0$\,\msun, $E_{\rm k}=6.6\times10^{50}$\,erg, and $\rho_{\rm CSM}=7\times10^{-13}$\,g\,cm$^{-3}$. Although this gives a satisfactory fit to the light curve, our spectroscopic modelling (Sect.~\ref{sec:synthesis_spec}) suggests that most of the luminosity is emitted by rapidly expanding material, therefore we disfavour this model relative to the magnetar scenario. However, as we discuss below, there may still be a (subdominant) contribution to the light curve from interaction.

Finally, we show a bolometric lightcurve model powered by the nuclear decay of $^{56}$Ni in Fig.~\ref{fig:sn_bol_model}. The $^{56}$Ni-powered LC model is derived in the same way as in \citet{2013ApJ...770..128I}. The model has a $^{56}$Ni mass of $7~\msun$ with $M_\mathrm{ej}=8~\msun$ and $E_\mathrm{k}=5\times 10^{52}~\mathrm{erg}$. The unrealistically high fraction of  $^{56}$Ni to ejecta disfavours the $^{56}$Ni-powered model. 
Fig.~\ref{fig:sn_bol_model} shows the magnetar model can mimic the $^{56}$Ni-powered lightcurve very well. 
As discussed in \citet{2017ApJ...835..177M}, they found that most $^{56}$Ni-powered lightcurves can be reproduced by magnetars that require the magnetar spin-down to be by almost pure dipole radiation with the breaking index close to 3.

\subsection{Late-time spectrum of LSQ14mo compared to SN~1998bw}
\label{sec:sn_results_spec_98bw}

The late-time spectrum of LSQ14mo taken with VLT around +300d is dominated by the host galaxy. We estimated the detection limit of \Oi 6300\AA\ , the strongest feature of type Ic SNe in nebular phase, by comparing the spectrum of LSQ14mo to the spectrum of the bright broad-line type Ic SN~1998bw at +337d \citep{2001ApJ...555..900P}. We did this in two ways.

Firstly, we scaled the spectrum of SN~1998bw to match the photometry and placed it at the same distance as LSQ14mo. We compensated for the difference in phase according to the $^{56}$Co decay rate and $\gamma$-ray trapping of SN~1998bw. We then shifted 
the SN~1998bw spectrum to match the red part of the continuum in LSQ14mo, and estimated the detection limit of \Oi 6300\AA\ in the spectrum of LSQ14mo at +300d to be $1.4\times10^{-18}$ erg\,s$^{-1}$\,cm$^{-2}$\,\AA$^{-1}$. This is similar to the comparison routines employed in \citet{2016arXiv160401226I}.

\begin{figure}
\includegraphics[angle=0,width=\linewidth]{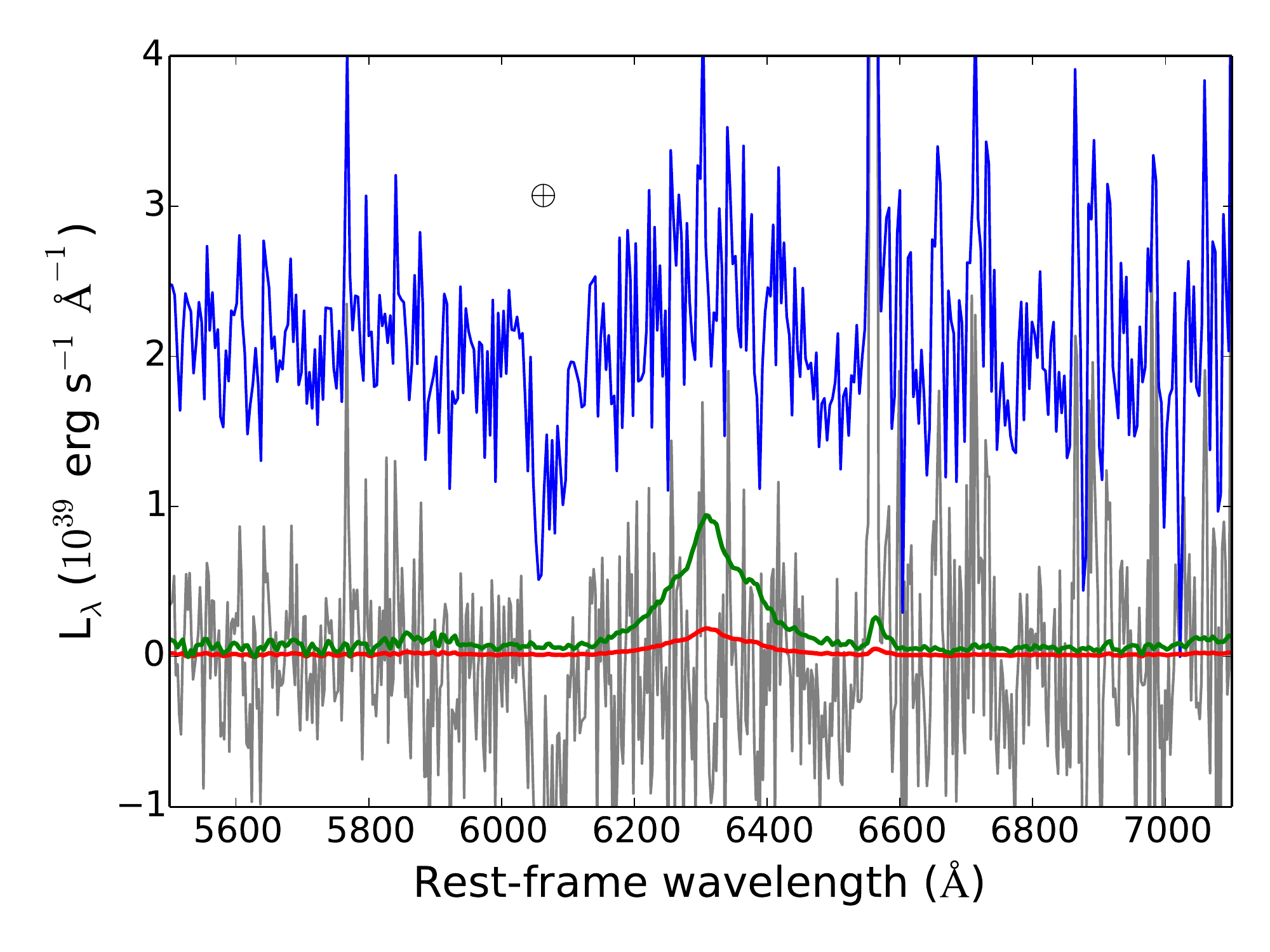}
\caption{VLT+FORS2 spectrum of LSQ14mo (position C) in black, with the stellar population model subtracted and  corrected to restframe luminosity. 
The restframe spectrum of the broad-line type Ic SN~1998bw at +337d \citep{2001ApJ...555..900P}, scaled to the distance of LSQ14mo, is shown in red, indicating the strength of 
the \Oi $\lambda\lambda$6300,6364 doublet. The green spectrum is SN~1998bw multiplied by 
a factor 5. The blue spectrum shows LSQ14mo with the SN~1998bw $\times$5 spectrum added, 
illustrating an upper limit to the detection of the \Oi $\lambda\lambda$6300,6364 doublet. 
A broad feature of this strength would have been visible in the LSQ14mo spectrum. The atmospheric telluric absorption features are marked with the earth symbol.}
\label{fig:14mo_98bw}
\end{figure}

Alternatively, we took the LSQ14mo FORS2 spectrum, subtracted the stellar population model 
at position C (see Fig.~\ref{fig:host_model_spec}), and corrected the spectrum to rest-frame luminosity 
($L_{\rm rest}(\lambda)=4\pi D_{\rm L}^{2}(1+z)f_{\rm obs}(\lambda)$). Then, we applied the same 
to the SN~1998bw spectrum, scaled it by five, and added it to the LSQ14mo spectrum. This
illustrates that we would have detected the \Oi 6300\AA\ doublet if it had been 
five times more luminous than in SN~1998bw, or if the luminosity was $L_{\rm 6300}\geq2.3\times10^{39}$ergs\,s$^{-1}$
(see Fig.\,\ref{fig:14mo_98bw}).

The peak bolometric luminosity of LSQ14mo is $10^{44.04}$\,erg\,s$^{-1}$, which is ten times brighter than that of SN~1998bw ($\sim 10^{43.05}$ erg\,s$^{-1}$).
\citet{2017ApJ...835...13J,2016ApJ...828L..18N} pointed out that the late-time spectrum of the SLSN SN~2015bn is very similar to SN~1998bw. 
Since our FORS2 spectrum shows that LSQ14mo is less than five times more luminous than SN~1998bw at late phases, this suggests an upper limit for the nickel mass of five times the nickel mass in SN~1998bw.
If we take 0.4\,\msun as the representative $^{56}$Ni mass for SN~1998bw \citep{2006ApJ...640..854M}, then the flux limit of LSQ14mo suggests an upper limit of        approximately 2\,\msun\ of $^{56}$Ni ejected by LSQ14mo. 
This is much lower than the 7\,\msun\ $^{56}$Ni mass needed to power the luminosity of LSQ14mo at maximum light (see Sect.\,\ref{sec:sn_results_bol_model}). Therefore, we rule out nickel-power as the underlying energy source in LSQ14mo. 
This spectroscopic late-time detection limit result supports the finding from \citet{2013ApJ...763L..28C}, who used a similar method. They found the upper limit of $^{56}$Ni production for the fast-declining SLSN SN~2010gx to be similar to or below that of SN~1998bw (0.4\,\msun). 
This small amount of $^{56}$Ni cannot reproduce such bright peak luminosity. However, the late-time ($+253$\,d) photometric detection limit of LSQ14mo ($9.31 \times 10^{37}$ erg\,s$^{-1}$) is not deep enough to compare with that of SN~1998bw ($1.35 \times 10^{37}$ erg\,s$^{-1}$).

\section{Synthetic spectrum modelling of LSQ14mo}
\label{sec:synthesis_spec}

\subsection{Theoretical modelling}
\label{sec:synthesis_spec_theory}

The spectra of LSQ14mo were modelled following the approach described in
\citet{2016MNRAS.458.3455M}. These models use a Monte Carlo supernova radiative transfer code (\citealt{1993A&A...279..447M, 2000ApJ...545..407M, 2000MNRAS.318...92L}), and have been used for SNe\,Ia \citep[e.g.][]{2014MNRAS.439.1959M} as
well as SNe\,Ib/c \citep{2013MNRAS.432.2463M}, and have recently been applied to SLSNe
\citep{2016MNRAS.458.3455M}\footnote{The luminosity of PTF09cnd is one order of magnitude smaller than what indicated in their Table 3. This error only affect the table, not the actual calculation.}

The code assumes that the SN luminosity is emitted at a sharp inner boundary and follows the propagation of energy packets through the expanding SN envelope, which has a characteristic SN density distribution. In the case of SLSNe, \citet{2016MNRAS.458.3455M} find that a steep density gradient yields good fits to the absorption-line spectra of different SLSNe. They find that thermal conditions in SLSNe explain some of the lines, but {\ion{O}{ii}} lines near maximum and \Hei lines after maximum require non-thermal excitation/ionisation, which they claim may be the result of X-rays leaking from a magnetar-interaction.

\subsection{Modelling of LSQ14mo}
\label{sec:synthesis_spec_14mo}

\begin{figure*}
\includegraphics[angle=0,width=\linewidth]{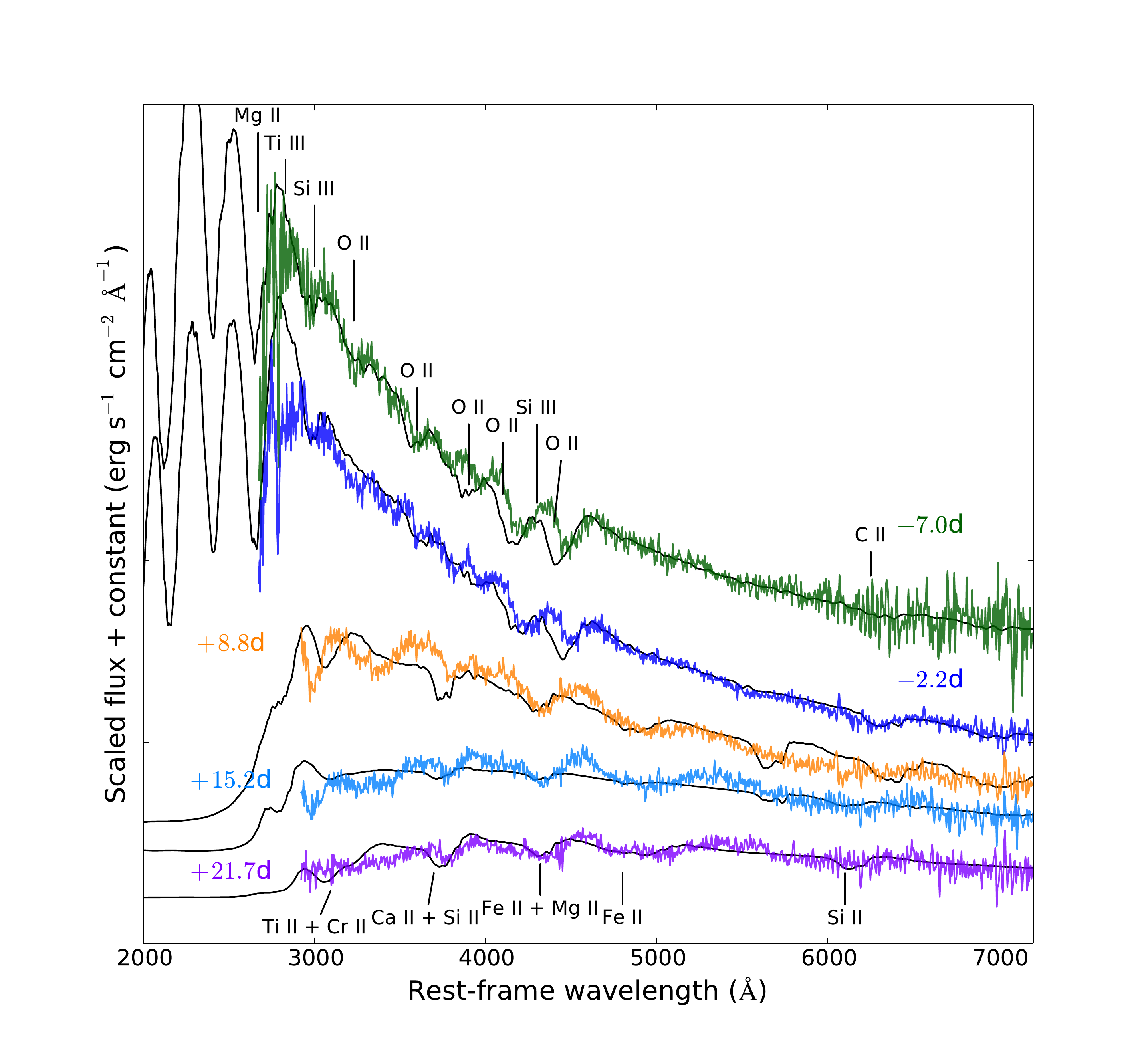}
\caption{We applied the spectral synthesis model (in black) from \cite{2016MNRAS.458.3455M} for LSQ14mo spectra. Each spectrum is labelled with its rest-frame time from the $r$-band lightcurve maximum.
The $-7.0$\,d spectrum is dominated by O and C, other intermediate-mass elements also shown and labelled. The $-2.2$\,d spectrum is similar to $-7.0$\,d in terms of element identification, temperature, luminosity and velocity. The  $+8.8$\,d spectrum, with a poor model fitting, shows a hot continuum but rather cool lines. This implies an additional luminosity may come from the SN ejecta interacting with a H- and He-poor shell. The $+15.2$\,d spectrum is similar to $+21.7$\,d and the strongest features are intermediate elements such as {\ion{Si}{ii}}, {\ion{Fe}{ii}}, {\ion{Mg}{ii}} and {\ion{Ca}{ii}}. All identified lines are labelled. }
\label{fig:14mo_synthesis_fit}
\end{figure*}

We adopt a model similar to those used in \citet{2016MNRAS.458.3455M}. We prefer a modest mass for the ejecta, $\sim 6$\,\msun. The kinetic energy is proportionally high, $\sim 7 \times
10^{51}$\,ergs, which gives a E/M (kinetic energy in units of $10^{51}$\,ergs and ejecta mass in $\log$\,\msun) ratio of $\sim 1$. 
This mass is slightly higher than, but similar to, that of our magnetar lightcurve fit of 4\,\msun (Sect. \ref{sec:sn_results_bol_model}), and the E/M ratio is also similar. Overall, the scenario we consider here is not inconsistent with the magnetar model inferred by the lightcurve fitting.
The spectral synthesis models for the LSQ14mo spectral time series are presented in Fig.\,\ref{fig:14mo_synthesis_fit} and discussed individually below.

The first spectrum at $-7.0$\,d has formally an epoch of 18 days after explosion (fits obtained from the model). This is well within the reconstruction of the detection limit and lightcurve of LSQ14mo. 
The $-7.0$\,d spectrum is characterised by a steep blue continuum and absorption lines of 
mostly {\ion{O}{ii}}. 
This spectrum has a luminosity $\log L$ = 44.10 (erg\,s$^{-1}$) and a photospheric velocity of 17000 km\,s$^{-1}$. 
The composition is (relatively) dominated by O (0.51) and C (0.31). We assume He (0.10) and lower abundances of Mg (0.01), Si (0.04), Ne (0.02) and other intermediate-mass elements,
including S, Ca, Ti and Cr. The Fe abundance 
in the layers above the photosphere is approximately 1/3 solar, consistent with the host metallicity measurement.
The strongest features reproduced are the {\ion{O}{ii}} near 3200, 3600, and 3900\AA\ and the 
distinctive 4200\,\AA\ (blended with {\ion{Si}{iii}}) and 4400\,\AA\ blends of {\ion{O}{ii}} 
transitions. The spectrum does appear to turn over at 3200, which is probably the 
{\ion{Si}{iii}} line near 3000\AA. Other very strong lines are predicted in the UV (further blue than our rest-frame spectra) including {\ion{Mg}{ii}} near 2600\,\AA.  The C\,{\sc ii} line at 6300 \AA\ is predicted
to be shallow and weak, and is not visible at the signal-to-noise of our spectra in this region.

The second spectrum at $-2.2$\,d has $\log L$ = 44.04 (erg\,s$^{-1}$) and a photospheric velocity of 15000 km\,s$^{-1}$. This is a small change in $L$ but a 2000\,km\,s$^{-1}$ change in velocity over five days. The observed spectrum is very similar to that of $-7.0$\,d. 
We used a very similar composition as for $-7.0$\,d, with slightly less Mg. The abundances of
several elements would be best set if the flux further in the UV was known,
which is unfortunately not the case. Our model reproduces the overall spectral
shape and the quality and strength of most lines. 
There may be reprocessed flux from the UV via metal lines in the observed spectrum, but it is unlikely that line fluorescence processes lead to such a smooth enhancement of the flux in such a limited spectral region. We may instead be witnessing the beginning of a short-lived additional energy source, as we discuss in the following section. 

The third spectrum, taken at $+8.8$\,d, is less blue and the {\ion{O}{ii}} lines have disappeared. 
This intermediate spectrum has a luminosity $\log$L = 43.83
(erg\,s$^{-1}$) and a photospheric velocity of 13750 km\,s$^{-1}$.
Other lines in the model are similar to the later epoch, but {\ion{Fe}{ii}} and {\ion{Mg}{ii}} are weaker. {\ion{Fe}{iii}} lines
are still present at 4700-5000\,\AA. 
The line at 4300\,\AA\ could be matched  with \Hei. Near 3700\,\AA,
{\ion{Ca}{ii}} is also very weak, and the line may be fit with \Hei. 
However, the fit to this spectrum is relatively poor, and some of the strongest predicted helium lines are not observed (e.g. 5876 and 6678\,\AA). 
 Therefore, the assumption of 10\% He may be too high, suggesting a highly-stripped core. It appears that while the luminosity is still high and the continuum blue ($T_{\rm eff} \simeq 10,000$\,K), most lines are compatible with a cool spectrum such as that observed at $+21.7$\,d. The spectrum of LSQ14mo changes quite dramatically between the previous epoch ($-2.2$\,d) and this one at $+8.8$\,d.

The fourth spectrum at $+15.2$\,d is very similar to that at $+21.7$\,d, but it also
has similar lines as that at $+8.8$\,d. Our model has $\log$L = 43.68
(erg\,s$^{-1}$) and a photospheric velocity of 13250 km\,s$^{-1}$. This means that a significant drop in luminosity is not accompanied by a drop in velocity. The model spectrum is redder than that at $+8.8$\,d, but it still does not reproduce the
observed lines very well because the temperature is too high. The composition
is the same as at $+21.7$\,d. Excess flux seems to be present from the NUV all the way to $\sim 6000$\,\AA. 

The fifth spectrum at $+21.7$\,d is relatively red, and it shows similar lines to the ones in the $+8.8$\,d spectrum.
It has a luminosity $\log$L = 43.53 (erg\,s$^{-1}$) and a
photospheric velocity of 12500 km\,s$^{-1}$. The composition is similar to the earlier epoch. The strongest features in the model are {\ion{Si}{ii}} near 6100\,\AA\ (a noisy region in the observed spectrum, but not clearly observed), {\ion{Fe}{ii}} multiplet 48 lines near 4700-4800\,\AA, {\ion{Fe}{ii}} and {\ion{Mg}{ii}} near 4300\,\AA, {\ion{Ca}{ii} blended} with {\ion{Si}{ii}} near 3700\,\AA, and {\ion{Ti}{ii}} and {\ion{Cr}{ii}} near 3100\,\AA. Heavy line blocking suppresses the NUV flux. Compared to iPTF13ajg \citep[][Fig. 4]{2016MNRAS.458.3455M}, the luminosity is a lot lower, singly ionised species are now present, and many of the features that were attributed to \Hei for iPTF13ajg spectra are fit by {\ion{Fe}{ii}} or {\ion{Si}{ii}} in LSQ14mo.

It should be noted that LSQ14mo is significantly less luminous than iPTF13ajg, and that it evolves much more rapidly than both iPTF13ajg and PTF09cnd (see Fig.\,\ref{fig:sn_phot_rf}), and therefore, direct comparisons must be treated with caution.
Near maximum, the {\ion{O}{ii}}-rich spectrum of LSQ14mo may be explained by non-thermal excitation. The later absence of He lines may be interpreted either as an indication that the He abundance is small, or that non-thermal processes are no longer at play, which is less likely. The modelling of iPTF13ajg \citep{2016MNRAS.458.3455M} indicated the presence of \Hei lines, but these were not required in PTF09cnd. Overall, we do not see an obvious requirement for helium in LSQ14mo.

\subsection{Interaction contribution seen in spectra}

\begin{figure}
\includegraphics[angle=0,width=\linewidth]{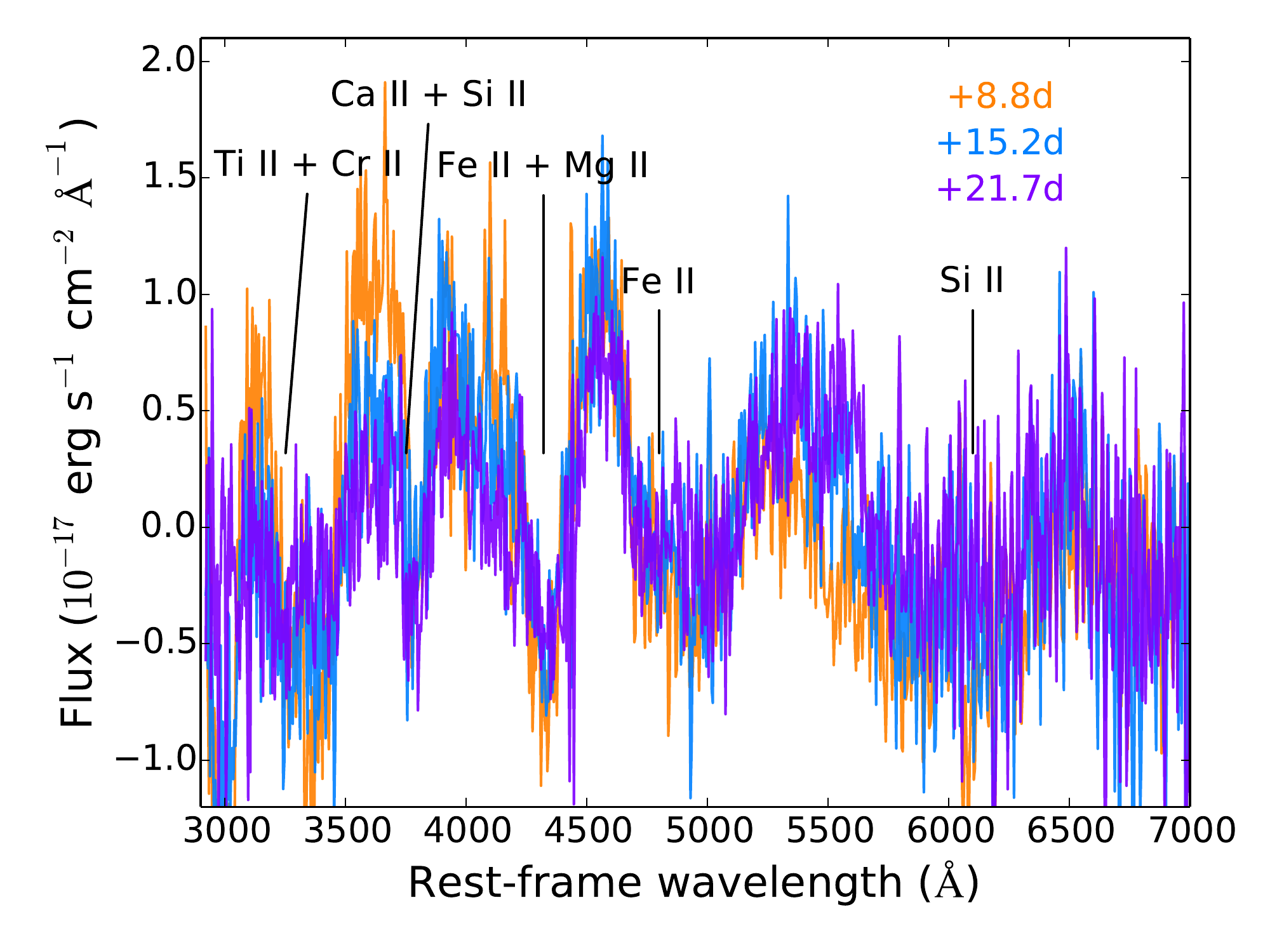}
\caption{Subtracted continuum (assumed black-body fitting) spectra of LSQ14mo at explosion epochs $+8.8$\,d, $+15.2$\,d, and $+21.7$\,d. The identification of spectral element features are the same at those three epochs. This supports the idea that the blue continuum observed in the $+8.8$\,d spectrum could be due to an additional energy source that likely comes from the interaction between the SN ejecta and a thin CO shell.}
\label{fig:14mo_spec_sub_BB}
\end{figure}

The spectrum after maximum ($+8.8$\,d) shows a hot continuum, but rather cool
lines. This is difficult to reproduce if the SN ejecta are in radiative
equilibrium, and is actually the opposite situation with respect to the epochs
when {\ion{O}{ii}} lines are present (in that case, the excitation temperature of the lines is above the continuum temperature). One option is that the SN absorption-line spectrum is indeed a cool one, and that additional luminosity is provided in a form that does not affect the lines. 
In Fig.\,\ref{fig:14mo_spec_sub_BB}, we present the spectra of LSQ14mo at $+8.8$\,d, $+15.2$\,d, and +$21.7$\,d with the continuum subtracted (assuming a black-body). The subtracted spectra have a similar profile; in fact, the line identifications are the same at those three epochs. 
This could arise if the ejecta were interacting with
some porous or clumpy, H- and He-poor shell, which might add a hot black-body component to the flux; for example, SN~2009dc had its luminosity augmented by interaction of the ejecta with a H-/He-poor circumstellar medium \citep{2012MNRAS.427.2057H}, as opposed to a H-rich one \citep[e.g.][] {2004ApJ...605L..37D}.

We suggest that we may be witnessing the contribution of a partially
thermalised, additional source of luminosity, starting after $-2.2$\,d and ending
before $+21.7$\,d. The spectrum at $+8.8$\,d is the most heavily affected. Because the
nature of the lines changes between $-2.2$\,d and $+8.8$\,d (from \ion{O}{ii} to \ion{Fe}{ii} and \ion{Mg}{ii}), while the continuum remains blue, it appears that
this additional component may not come from the same physical location where the lines are formed.

\begin{table*}
 \begin{minipage}{175mm}
 \centering
  \caption{Temperature, photospheric velocity and luminosity evolution of LSQ14mo from the spectral modelling results. The phase (day) has been corrected for time dilation ($z = 0.256$) and relative to the SN {\it r}-band maximum on MJD 56697. The explosion epoch is given from the spectral model. Temp$_{R}$ refers to radiation temperature and Temp$_{*}$ is effective temperature. Additionally, we list a black body temperature (Temp$_{BB}$) estimated from fitting the spectrum assuming a black body; this value is similar with the radiation temperature. $v_{5169}$ measured from minimum of \Feii $\lambda5169$ absorption is also shown.}
\label{tab:sn_temp_vel}
\begin{tabular}[t]{llllllllll}
\hline\hline 
Date & Phase & Exp. epoch & Temp$_{R}$ & Temp$_{*}$ &  $\mathit{log}$\,L & Velocity & Temp$_{BB}$ & $v_{5169}$\\
  & (day) & (day)  & (K) & (K) &  (erg\,s$^{-1}$) & (km\,s$^{-1}$)  & (K) &(km\,s$^{-1}$) \\
\hline
2014 Jan 31     & $-7.0$        &  $\mathit{18}$ & 14885 & 13272&44.10 & 17000& 14641 & - \\
2014 Feb 6      & $-2.2$        & $\mathit{23}$ & 13971 & 12351 &44.04 & 15000  & 13636 & -\\
2014 Feb 20 & +8.8      & $\mathit{34}$ & 9674 & 8961 &43.83 & 13750 &10083 & $10500\pm1500$\\
2014 Feb 28     & +15.2 & $\mathit{40}$ & 7698 &7465& 43.68 & 13250& 7421 & $10400\pm500$\\
2014 Mar 8      & +21.7 & $\mathit{47}$ & 6965 &6700&43.53 & 12500& 6488 & $10200\pm1500$\\
\hline
\end{tabular}
\end{minipage}
\end{table*}

\section{Host galaxy properties}
\label{sec:host_galaxy_properties}

\subsection{Galaxy size, extinction corrections, and luminosity}
\label{sec:galaxy_size}

The profile of the host galaxy is noticeably broader than the stellar PSF, hence we can determine the physical diameter of the extended source, assuming the relation (galaxy observed FWHM)$^{2}$ = (PSF FWHM)$^{2}$ + (intrinsic galaxy FWHM)$^{2}$. We measured the FWHM at both the host positions A (1.74\arcsec) and C (1.03\arcsec) and the average FWHM (0.80\arcsec) of seven reference stars within a 2\arcmin\ radius around the host on the NTT $i$-band image taken on 2015 February 12, which had the best seeing conditions (0.8\arcsec). We were unable to fit the FWHM of a Gaussian profile at position B because of its irregular profile.
This provides a physical diameter of 6.17\,kpc for position A at the angular size distance of 824.4\,Mpc, and a physical diameter of 2.63\,kpc for position C at 826.0\,Mpc. 
For comparison, the size of the LMC is $\sim 4.3$\,kpc in diameter, and 2.1\,kpc for the SMC. 

The angular separation between positions A and C is $\sim$ 4\arcsec, measured from the brightest region of each galaxy component,  
which corresponds to $\sim 15.2$\,kpc at the distance of the LSQ14mo host system.
The separation between positions A and B is $\sim10.5$\,kpc, and between positions B and C it is $\sim4.7$\,kpc. Those separations were consistent with that measured from the emission line positions in three spectral traces.

We used the Balmer decrements to estimate the internal dust extinction of the host galaxy system. The intrinsic line ratio is expected to be \ha/\hb = 2.86, assuming case B recombination for $T\rm_{e}$ = 10000\,K and $n \rm_{e}$ = 100\,cm$^{-3}$ \citep{1989agna.book.....O}. 
The measured fluxes (Table\,\ref{tab:galaxy_flux}) of \ha and \hb give a ratio of 3.10 at position A, which indicates an internal dust extinction of approximately {\it A$_{V} = 0.26$} in the rest-frame, assuming {\it R$_{V}$} = 3.1. For positions B and C, the \ha/\hb value is slightly lower than 2.86, corresponding to 2.57 and 2.74, respectively.
Therefore, we assumed that the internal dust reddening is negligible in positions B and C, that is,~{\it A$_{V}$} = 0.

Finally, we obtained the absolute $g$-band magnitudes ($M_{g}$) from the observed $r$-band magnitude ($m_{r}$) after applying {\it K}-correction (Table\,\ref{tab:sn_K-correction}), foreground and internal dust extinctions, using the formula:
$M_{g} = m_{r} - A_{r{\rm ,\,Milky Way}} + K_{r \rightarrow g} - A_{g{\rm ,\,host}}$.
This gives us {\it M$_{g}$} = $-19.04\pm0.15$\,mag, $-16.94\pm0.15$\,mag, and $-16.85\pm0.15$\,mag at the three components A, B, and C, respectively.

\begin{table*}
 \begin{minipage}{175mm}
 \centering
  \caption{Main properties of the host galaxy system of LSQ14mo. The SN site is located at the position C. }
\label{tab:galaxy_property}
\begin{tabular}[t]{llll}
\hline\hline 
& Position A & Position B & Position C \\
\hline
RA (J2000)& 10:22:41.55 & 10:22:41.52  & 10:22:41.48 \\ 
Dec (J2000)&  $-16$:55:17.9& $-16$:55:15.2  & $-16$:55:14.2  \\
Redshift & 0.2556 & 0.2553 & 0.2563\\   
Projected separation (kpc) & - & 10.5 (A to B) & 15.2 (A to C) \\
Physical diameter (kpc) & 6.17& - &2.63\\ 
Radial velocity offset (km\,s$^{-1}$) & - & -90 (A to B) & 210 (A to C) \\
Apparent $r$ (mag) & $21.26\pm0.02$ & $23.58\pm0.06$ & $23.60\pm0.06$\\
Galactic extinction $A_{V}$ (mag) & 0.20 &0.20 & 0.20\\
Internal extinction $A_{V}$ (mag) & $\sim$ 0.26 & $\sim$ 0.0 & $\sim$ 0.0\\
Absolute $g$ (mag) & $-19.04\pm0.15$& $-16.94\pm0.15$ & $-16.85\pm0.15$ \\
GALEX $NUV$ (mag) & $22.87 \pm 0.39$ &-&-  \\
H$\alpha$ Luminosity (erg\,s$^{-1}$) &$1.04\times10^{41}$&$1.62\times10^{40}$&$1.30\times10^{40}$ \\
H$\alpha$ EW$_{rest}$ (\AA) & 54.10 & 174.67 & 84.94\\
SFR from \ha ($M_{\odot}$\,yr$^{-1}$) &0.52&0.08&0.06\\
$\log$ Stellar mass ($M_{\odot}$) & $8.8^{+0.1}_{-0.2}$ & $7.6^{+0.3}_{-0.2}$ & $7.7^{+0.2}_{-0.2}$ \\
sSFR (Gyr$^{-1}$) & 0.82&1.89&1.16 \\
$12+\log {\rm(O/H)}$ (T{\rm$_{e}$}) & $8.00\pm0.31$ & - & - \\
$12+\log {\rm(O/H)}$ (KK04 $R_{23}$) & $8.62\pm0.03$& $8.28\pm0.03$& $8.18\pm0.02$ \\
$12+\log {\rm(O/H)}$ (PP04 N2) & $8.31\pm0.01$& $8.09\pm0.05$ & $8.18\pm0.05$\\
$12+\log {\rm(O/H)}$ (D16) & $8.02\pm0.03$&  - & $7.92\pm0.16$ \\
Youngest stellar population from \ha (Myr) & 6.3-9 & 6.3 &  7.7 \\
\hline 
\end{tabular}
\end{minipage}
\end{table*}

\subsection{Stellar population ages in the host galaxy system}
\label{sec:stellar_population}

There are multiple stellar populations inside the galaxy. We used the EW of \ha line as a tracer for ongoing star formation and as an age indicator to probe a young stellar population of \Hii region. Such methods have also been employed recently to analyse the spectra of the environments of local SNe \citep[e.g.][]{2011A&A...530A..95L, 2013AJ....146...30K}. 
We found the rest-frame EWs of \ha to be 54\AA, 175\AA,\ and 85\AA\ at positions A, B, and C, respectively. 
To be more consistent with our metallicity measurement of the host system of LSQ14mo, we chose the new simple stellar population (SSP) model with low-metallicity (0.2 \zsol) from \citet{2016A&A...593A..78K}, which gave slightly older ages than that given by solar metallicity. 
We further considered the SSP with a Kroupa IMF \citep{2013ApJ...779..170L}, which is similar to our assumption of a Chabrier IMF. Our EW range of the \ha line is much smaller than the continuous star-formation models ($> 500$\AA) and therefore we chose the SSP model of an instantaneous burst of star formation. 
This resulted in a stellar population age between 6.3 and 9\,Myr for position A (as its metallicity of 0.4-0.8 \zsol is between the SSP models of 0.2-1 \zsol), 6.3\,Myr at position B and 7.7\,Myr at position C. The youngest stellar population is located at the B position, while older, relatively similar stellar populations are present at positions A and C (SN location). 
The largest EW of \ha line at position B supports our interpretation that the position B is a new-born bright \Hii region from the interacting activities of galaxies A and C.

Galaxies certainly contain multiple stellar populations rather than a single one.
We also reported the $r$-band light-weighted stellar population age given by {\sc magphys}. The mean age of position A is slightly older at 212\,Myr, than the 94\,Myr and 118\,Myr at positions B and C, respectively. Alternatively, the fitting from spectral continuum using {\sc starlight} also implies both young (3-7\,Myr) and old (500\,Myr) populations at positions A and C. However, those values are model dependent and there are not only those populations inside the galaxy. 
We can only argue that a young stellar population is present as \ha emission line is detected. 
For comparison, to date, \citet{2015MNRAS.451L..65T} reported the largest EW of \ha line ($>$ 800\AA) among the SLSN host galaxies that has been seen in the host of PTF12dam, and thus suggested a very young stellar population at the SN site of $\sim 3$\,Myr.

\subsection{Stellar masses and star-formation rates in the host galaxy system}
\label{sec:stellar_mass_sfr}

\begin{figure}
\includegraphics[angle=0,width=\linewidth]{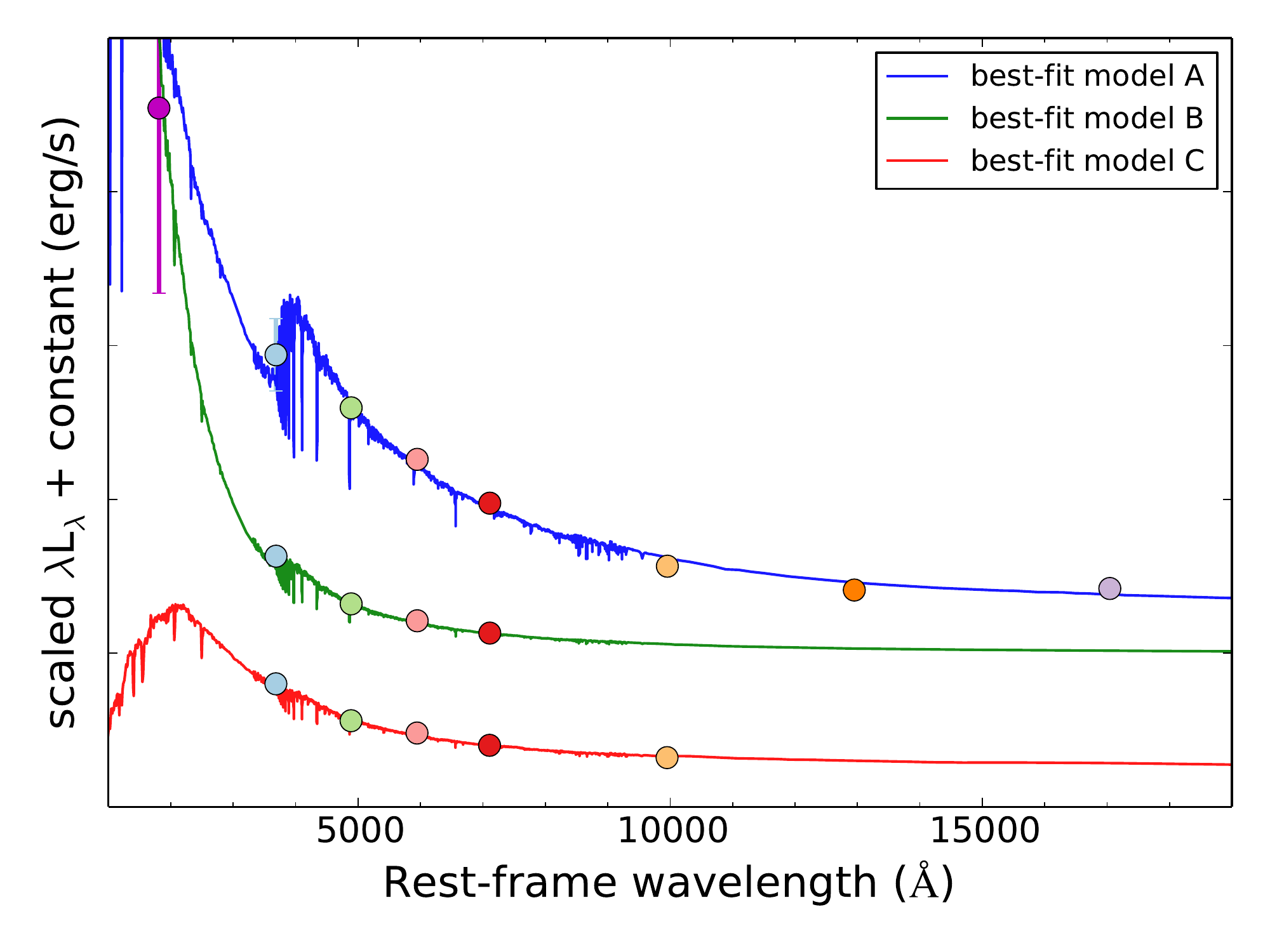}
\caption{The best-fit SED models of the host system of LSQ14mo from {\sc magphys}, for three components, A, B, and C, are shown in blue, green, and red lines. 
The colour circles show the different bands' photometry (shifted to the rest-frame) for the three components of host system where it was possible to have a clear detection. For instance, for component A, the colour circles from left to right are from $NUV$, $griz$ to $JHK$-bands photometry.
The median and a $1\sigma$ range of stellar masses of components A, B, and C are: $8.8^{+0.1}_{-0.2}$, $7.6^{+0.3}_{-0.2}$ and $7.7^{+0.2}_{-0.2}$\,$\log$M/\msun, respectively. 
}
\label{fig:host_sed_mass}
\end{figure}

To obtain stellar masses of the LSQ14mo host galaxy system, we fit the available photometry with the stellar population models in {\sc magphys} \citep{2008MNRAS.388.1595D} using \citet{2003MNRAS.344.1000B} templates.  The best fit is shown in Fig.\,\ref{fig:host_sed_mass}, and returns masses (in unit of $\log$M/\msun) of $8.8^{+0.1}_{-0.2}$ for position A, $7.6^{+0.3}_{-0.2}$ for position B, and $7.7^{+0.2}_{-0.2}$ for position C. 
Here, the stellar mass is the median of the probability density function (PDF) over a range of models and the errors correspond to the $1 \sigma$ credible intervals of the PDF. For comparison, the stellar mass of the SN position (C) given by \citet{2016arXiv161205978S} of $7.89^{+0.15}_{-0.19}$\,$\log$M/\msun from the SED fitting is shown, which is consistent with our estimate.

After correcting the \ha line luminosity for Milky Way and internal host dust reddening, we applied  the conversion of ${\rm SFR = L(\ha)} \times 7.9 \times 10^{-42}$ to calculate the star-formation rate \citep[SFR;][]{1998ARA&A..36..189K}, and then divided it by 1.6 assuming a Chabrier IMF. 
The SFR  is 0.52 \msun yr$^{-1}$for component A, 0.08 \msun yr$^{-1}$ for component B and 0.06 \msun yr$^{-1}$ for component C. 
Together with their stellar masses, we determined a specific SFR (sSFR) of 0.82\,Gyr$^{-1}$, 1.89\,Gyr$^{-1}$ , and 1.16\,Gyr$^{-1}$ for components A, B, and C, respectively.

\subsection{Gas-phase metallicity of the host galaxy}
\label{sec:metallicity}

We carry out an abundance analysis based on both the strong line methods and the ``direct'' measurement, derived from the weak \Oiii $\lambda4363$ auroral line detected at position A. The emission line fluxes (Table\,\ref{tab:galaxy_flux}) have been corrected for Milky Way and internal dust extinction (Sect.\,\ref{sec:galaxy_size}).  

First of all, we checked the locations of the host system of LSQ14mo on the BPT diagram \citep{1981PASP...93....5B}. 
The $\log$(\Oiii/\hb) = 0.39, 0.58, and 0.52 and $\log$(\Nii/\ha) = $-0.91$, $-1.47,$ and $-1.24$ for positions A, B, and C, separately. 
They are all located at the star-forming galaxy region, while components B and C are near the high ionised boundary with $\log$(\Oiii/\hb) $> 0.5$. This characteristic of SLSN host galaxies has been noticed in \citet{2015MNRAS.449..917L}, and the host galaxy of LSQ14mo (position C) also follows this trend. 

We used the open-source {\sc python} code {\sc pymcz} \citep{2016A&C....16...54B} to calculate the oxygen abundance in different diagnostics. We adopted the \citet{2004ApJ...617..240K} (hereafter KK04) calibration of the $R_{23}$ method, which uses the (\Oiii $\lambda\lambda4959,5007$ + \Oii $\lambda3727$)/\Hb{} ratio. 
We found an oxygen abundance of $12 + \log({\rm O/H}) = 8.62 \pm 0.03$, $8.28 \pm 0.03,$ and $8.18 \pm 0.02$ for positions A, B, and C, respectively (where the H$\alpha$ and \Nii$\lambda$6584 ratio was used to break the $R_{23}$ degeneracy). We also used the N2 method of \citeauthor{2004MNRAS.348L..59P} (2004, hereafter PP04), which uses the log(\Nii$\lambda6583$/\ha) ratio, giving 12 + $\log$(O/H) = $8.31 \pm 0.01$, $8.09 \pm 0.05,$ and $8.18 \pm 0.05$ for positions A, B, and C, respectively. 
In addition, we considered a new metallicity diagnostic proposed by \citeauthor{2016Ap&SS.361...61D} (2016, hereafter D16), which uses \Nii$\lambda6583$, \ha, and \Sii$\lambda\lambda6717,6731$ lines.
We found 12 + $\log$(O/H) = $8.02 \pm 0.03$ for position A and  $7.92\pm0.16$ for position C.
We note that, for position B, the line ratio combination required for the D16 diagnostic returns a value ($y = -1.12$) at the edge of where their calibration is applicable. Therefore, we did not apply this metallicity diagnostic at position B.

We also attempted an estimate of the metallicity at position A from direct measurements of the \Oiii electron temperature. The auroral \Oiii $\lambda4363$ line is intrinsically weak and difficult to detect since all SLSNe have been found at distances greater than $z = 0.1$ \citep[see][]{2015MNRAS.452.3869N}. The marginal detection (S/N = 3.6) of the auroral line seen in the spectrum of position A gives an electron temperature of $T_{{\rm e}}= 13000$\,K assuming an electron density of 100 cm$^{-3}$. This indicates a metallicity of 12 + $\log$(O/H) = $8.00 \pm 0.31$.

The metallicities obtained from each of these diagnostics are reported in Table.\,\ref{tab:galaxy_property}.
To conclude, the SN location (position C) exhibits the lowest metallicity when using the KK04 and D16 diagostics, even when accounting for the $\sim 0.1$ dex systematic uncertainty in these diagnostics. This is perhaps unsurprising, given the know correlation between $M_{*}$ and metallicity (\eg \citealt{2004ApJ...613..898T}), as position C also has the lowest stellar mass.
It is known that there is a systematic offset of 0.2-0.4 dex (and sometimes up to 0.6 dex) between the KK04 method and both the PP04 and direct $T_{\rm e}$ method \citep{2007A&A...473..411L, 2011ApJ...729...56B}, with the former giving systematically higher values. As discussed in \citet{2008ApJ...681.1183K}, for example, position A with a stellar mass of 8.80 $\log$M/\msun, a typical offset between KK04 $R_{23}$ and PP04 N2 is 0.35 dex, which is consistent with the values we measured.

We further obtained a nitrogen-to-oxygen ratio of $\log{\rm (N/O)} = -1.35 \pm 0.24$ at position A from the $R_{23}$ calibration method, assuming that N/O = N$^{+}$/O$^{+}$. We measured $\log{\rm (N/O)} = -1.58 \pm 0.86$ at position B, and $-1.36 \pm 1.60$ at position C. 
This places those three components close to the ``plateau'' seen at $\log({\rm N/O}) \simeq -1.5$ on the N/O versus O/H diagram of galaxies \citep{2003A&A...397..487P}. For comparison, the mean values of oxygen and nitrogen abundances in 21 \Hii regions of the Small Magellanic Cloud are $12 + \log({\rm O/H}) = 8.07\pm0.07$ ($T_{\rm e}$ method) and $\log({\rm N/O}) =-1.55\pm0.08$ 
\citep[from literature values summarised in][]{2003A&A...397..487P}. Hence, the metallicity environments in the host system of LSQ14mo are similar to (or perhaps somewhat more extreme than) that of the SMC.

\section{Discussion I: superluminous supernovae with interacting features}
\label{sec:discussion_sn}

As the hot continuum has been seen in our $+8.8$\,d spectrum, we propose that this additional energy is from the interaction between the SN ejecta and a thin shell or some clumps inside the CSM. This material likely originates from a mass loss before the SN explosion. Here we roughly calculated an upper limit on the energy input from such as interaction, and the place of this interacting shell.

\begin{figure}
\includegraphics[angle=0,width=\linewidth]{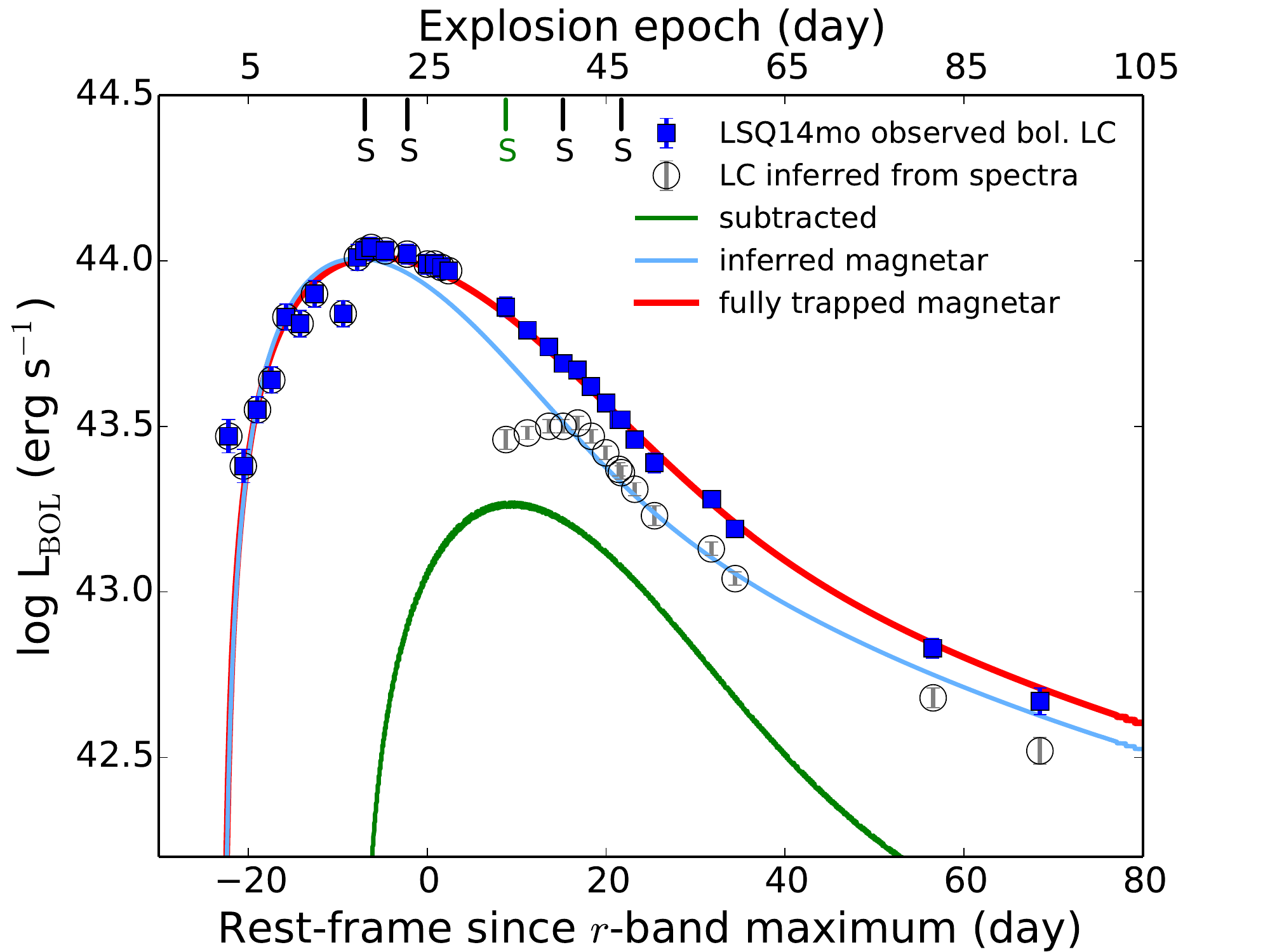}
\caption{Combination of magnetar-powered and interaction contribution of bolometric luminosity of LSQ14mo. The symbol `S' marks the time when spectra were taken, and the green `S' highlights the epoch with a hot continuum seen in the spectrum. Blue squares show the final bolometric lightcurve integrated from $UVW2$ to $K$ and fitted with a fully trapped magnetar model in red; black empty circles show the lower limit of assumed magnetar-powered lightcurve and fitted in blue. The luminosity difference between red and blue magnetar models is an additional energy source in green, which  is
speculated to come from the interaction between the SN ejecta and a thin shell or some clumps inside the CSM likely originating from a massive star mass loss before the SN explosion.}
\label{fig:14mo_bol_interaction}
\end{figure}

Firstly, we assumed $+21.7$\,d spectrum as a baseline, and that any additional luminosities at $+8.8$\,d and $+15.2$\,d spectra came from the interaction. This is of course an upper limit on the interaction energy, since the SN ejecta also cools down from $+8.8$\,d to $+21.7$\,d. The upper limit on the luminosity from the interaction is up to 60\% at $+8.8$\,d and 36\% at $+15.2$\,d. From our synthetic spectrum modelling results, the two early-phase spectra ($-7.0$\,d and $-2.2$\,d) were well fitted by the model (see Fig.\,\ref{fig:14mo_synthesis_fit}). Hence we propose  that the interaction only happened after $-2.2$\,d and that its energy contribution peaked at $+8.8$\,d. We assumed that the interaction plays a role of 60\% at $+8.8$\,d and 36\% at $+15.2$\,d, and keep 30\% after $+15.2$\,d while 0\% around and before the SN peak. This is clearly a coarse assumption, but we use this simple approach to illustrate the essential points. Fig.\,\ref{fig:14mo_bol_interaction} demonstrates the assumed bolometric lightcurve after subtracting the interaction contribution. We assumed the remaining luminosity is powered purely by a magnetar. 
Fitting this light curve (reduced $\chi^{2} = 9.53$), we obtain a magnetic field strength of $5.5\times 10^{14}~\mathrm{G}$ and an initial spin period of 4.8~ms with $M_{\rm ej} = 2.7~\msun$.
We subtract this new magnetar model fit from the model for the full bolometric light curve (Sect.~\ref{sec:sn_results_bol_model}) and plot the difference in Fig.\,\ref{fig:14mo_bol_interaction} to give an estimate of the light curve component from interaction.

Secondly, we estimated the location of this interacting material. We determined an initial radius ($R_{0}$) of the SN photosphere at $-7.0$\,d from the equation of $L = 4 \pi R_{0}^{2} \sigma T^{4}$, where $\sigma$ is the Stefan-Boltzmann constant, and we used the effective temperature and luminosity listed in Table\,\ref{tab:sn_temp_vel}. Thus we got the $R_{0} = 2.39 \times 10^{15}$ cm.
We approximated the radius of the next epoch to $R = R_{0} + v \times \delta_{t}$, where $v$ is the photospheric velocity, which we adopted to be the average between two epochs from $R_{0}$ to $R$. Then we moved on to the next epoch of $+8.8$\,d, and obtained the radius of $4.51 \times 10^{15}$ cm of this interacting thin shell.

\begin{figure}
\includegraphics[angle=0,width=\linewidth]{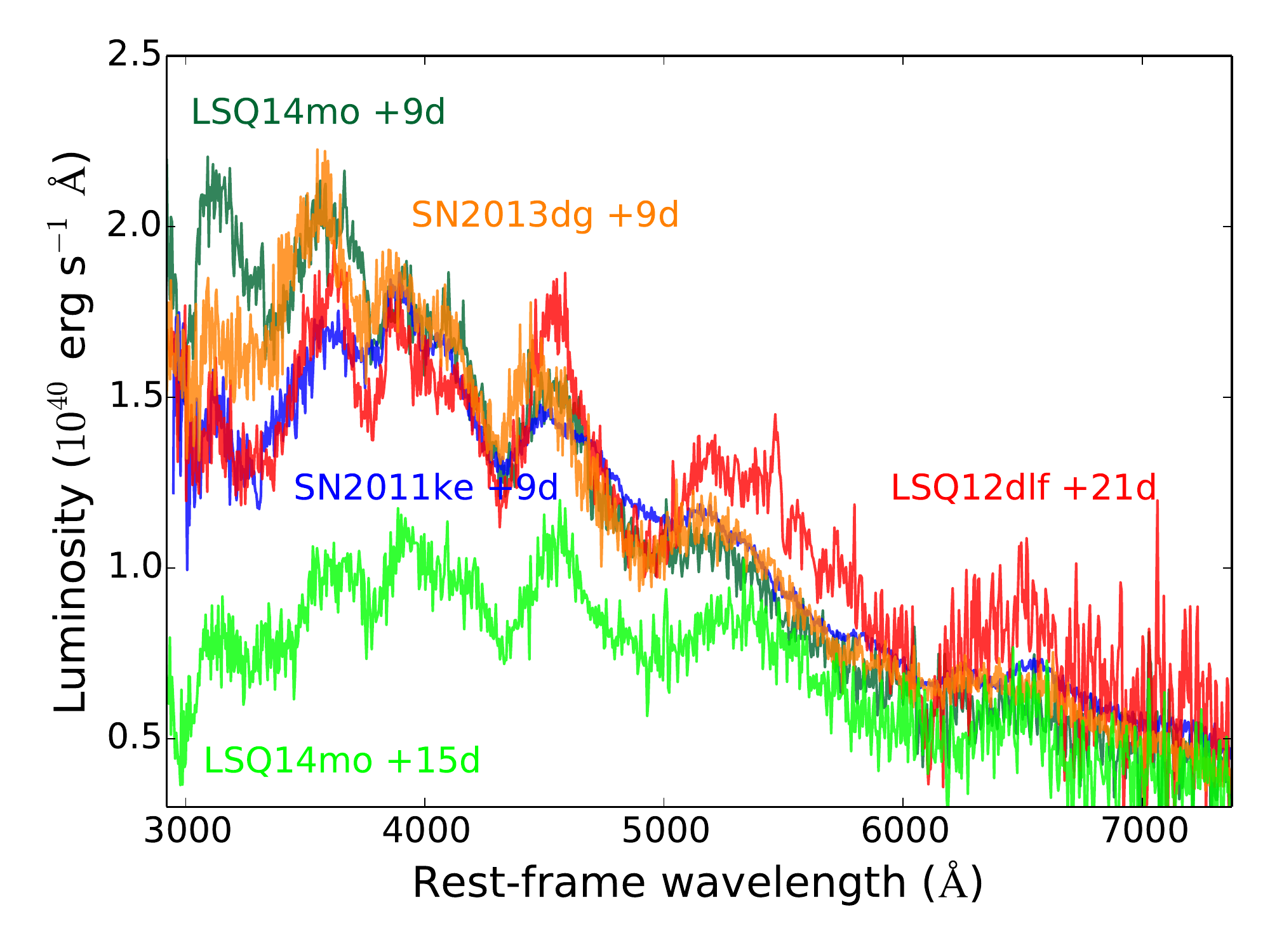}
\caption{Spectral comparison between LSQ14mo and other fast-declining SLSNe at a similar observational phase. The interacting feature of a blue continuum but rather cool lines is also shown in SN~2013dg at +9d, SN~2011ke at +9d, and LSQ12dlf at +21d. Data for comparison are taken from \citet{2013ApJ...770..128I,2014MNRAS.444.2096N}.
}
\label{fig:14mo_spec_com_10d}
\end{figure}

LSQ14mo is the first event for which interacting features in fast-evolving SLSN spectra have been reported, by seeing a hotter continuum and cooler lines.  
We therefore searched for other SLSNe with similar behaviour and found several fast-declining SLSNe at a similar phase to LSQ14mo, with interaction around 10 days after the maximum light. They are:
SN~2013dg at +9\,d, SN~2011ke at +9\,d, and LSQ12dlf at +21\,d, shown in Fig.\,\ref{fig:14mo_spec_com_10d}.
A further larger sample study will be interesting to investigate.

Overall, LSQ14mo is one of the fastest evolving SLSNe of type I which has a well constrained rise and decline (see Fig.\ref{fig:sn_phot_rf}). This fact leads to a relatively low ejecta mass of approximately 4\,\msun 
from the magnetar powering fits. This mass is on the lower end of the distribution of low redshift 
SLSNe Ic presented by 
\cite{2015MNRAS.452.3869N}. For such low masses, magnetar powering inherently implies formation of a neutron star remnant and hence a progenitor mass of $\sim$6\msun. Such a carbon-oxygen star would be relatively low in mass for the final mass of a Wolf-Rayet star \citep{2007ARA&A..45..177C}.

\section{Discussions II: Interacting host galaxy}
\label{sec:discussion_host}

\subsection{Star-formation activity}
\label{sec:dis_SF_activity}

\begin{figure*}
 \includegraphics[angle=0,width=\linewidth]{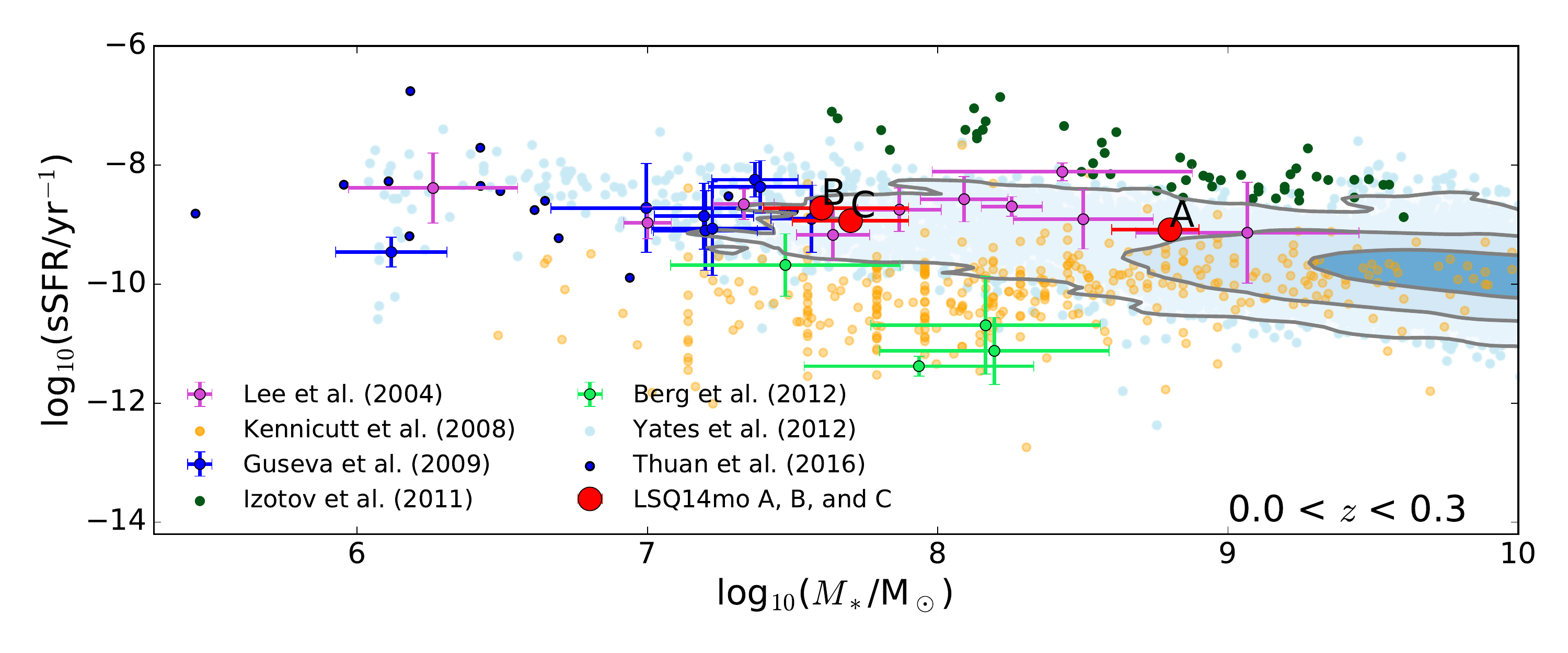}
 \caption{Relation between stellar mass ($M_{*}$) and specific SFR (sSFR) for positions A, B, and C (red points) from LSQ14mo system compared to a selection of local dwarf galaxies ($z < 0.3$). The contours represent the star-forming sample of \citet{Yates+12}, showing the 68th, 95th, and 99.5th percentiles of the distribution. All three LSQ14mo systems have SFRs (and sSFRs) slightly higher than most galaxies of their mass.}
\label{fig:Mstar-sSFR}
\end{figure*}

In Fig. \ref{fig:Mstar-sSFR}, we plot the relation between $M_{*}$ and specific SFR (sSFR = SFR/$M_{*}$) for systems A, B, and C, along with a selection of local galaxies. SFRs for the \citet{Lee+04}, \citet{Kennicutt+08}, \citet{Izotov+11}, and \citet{Thuan+16} samples are measured via the dust-corrected H$\alpha$ flux or narrow-band luminosity, using the $L(H\alpha)$-SFR relation from \citet{1998ARA&A..36..189K} and the \citet{1989ApJ...345..245C} extinction law. Total SFRs for the \citet{Guseva+09}, \citet{Yates+12}, and \citet{Berg+12} samples were obtained from the SDSS-DR7 spectroscopic catalogue \citep{Brinchmann+04}. All SFRs and stellar masses have been corrected to a Chabrier IMF.

We note that the \citet{Kennicutt+08} and \citet{Berg+12} data (orange and green points) contain very local galaxies ($z < 0.002$), the \citet{Izotov+11} data (dark green points) comprise `green pea' galaxies with particularly high specific SFRs, and the other data represent a statistically significant sample of H$\alpha$-bright dwarf galaxies out to a redshift of $z = 0.3$. Therefore, these samples combined constitute a fair representation of the range of sSFR found in the local dwarf galaxy population.

We can see from Fig. \ref{fig:Mstar-sSFR} that the three LSQ14mo systems each have typical sSFR compared to the general star-forming dwarf-galaxy population out to $z = 0.3$. This is consistent with the idea that SLSN hosts, such as system C, need to have a high sSFR in an absolute sense [\ie log(sSFR$/\textnormal{yr}^{-1})\gtrsim -10.3$] \citep{2015MNRAS.449..917L, 2016arXiv160504925C}, but not necessarily higher than other galaxies of the same stellar mass.

\subsection{Interacting galaxy system} 
\label{sec:dis_Interacting_system}

All three systems, A, B, and C, are in close proximity to each other. Their projected separations are $r\sub{proj,AB}\sim 10.5$ kpc, $r\sub{proj,AC}\sim 15.2$ kpc, and $r\sub{proj,BC}\sim 4.7$ kpc, and their line-of-sight velocity ($v\sub{los} \equiv cz$) differences are $\Delta{}v\sub{los,AB} = 89.94$ km/s, $\Delta{}v\sub{los,AC} = -209.86$ km/s, and $\Delta{}v\sub{los,BC} = -299.80$ km/s, respectively. This means that all three systems are easily close enough to be considered interacting under typical selection criteria (\eg \citealt{Kewley+06, Ellison+08, Michel-Dansac+08}).

Interacting systems are known to exhibit enhanced SFRs and lowered gas-phase metallicities compared to isolated galaxies of the same mass (\eg \citealt{Lee+04, Peeples+09}). This is also seen for other SLSN type I host galaxies with disturbed morphologies (\citealt{2015ApJ...804...90L, 2016ApJ...830...13P}). 
However, none of the three systems in LSQ14mo have particularly high sSFR or low $Z\sub{g}$ for their stellar mass. Integral field unit (IFU) observations are therefore required to investigate this host system more carefully, in order to better determine its interacting state.

The maximum escape velocity of system C from A (\ie assuming that their true separation is simply their projected separation, $r\sub{proj,AC}$) can also be estimated. Firstly, assuming system A is a central galaxy, we can use the multi-epoch abundance matching technique of \citet{Moster+13} to obtain an estimate of its dark matter (DM) halo mass. Abundance matching statistically determines the relationship between stellar mass and DM halo mass by fitting the observed galaxy stellar mass function to simulations of dark matter halo growth (\eg \citealt{Moster+10}). For the stellar mass and redshift of system A, this technique estimates a DM halo mass of $\textnormal{log}(M\sub{DM}/\msun) \sim 11.0$. In order to estimate the amount of this DM that is within $r\sub{proj,AC}$, we further assume the standard NFW \citep{Navarro+97} DM density profile and a DM concentration parameter of 16, both of which are suitable for DM haloes of this size \citep{Sawala+15,Bullock+01}. This implies that $\sim 29$ per cent of the DM is within the 15.2 kpc projected radius of system C, and that the maximum escape velocity of system C from A is $v\sub{esc} = (2GM\sub{tot}/r\sub{proj,AC})^{1/2} \sim 130$ km/s. This is only $\sim 62$ per cent of the (orthogonal) relative line-of-sight velocity of $\Delta{}v\sub{los,AC} = -209.86$ km/s, suggesting that the system {may} not be gravitationally bound and that system C could be in the process of a fly-by past system A.

System B has the largest \ha line EW, indicating that it contains the youngest stellar population. In addition, its weak stellar continuum suggests that system B is not a galaxy but a new-born bright \Hii region, possibly formed due to the interaction between galaxies A and C. This is similar to the Antennae galaxies (NGC 4038/NGC 4039) where many \Hii regions with high star-formation are distributed around the off-nuclear regions \citep{2004ApJS..154..193W}. 
This kind of starbursting clump is also seen in other galaxy merger systems. For example, in tidal features and very young stellar populations (\eg \citealt{2009AJ....137.4643H, 2013ApJ...779..170L}).

\subsection{The stellar population of complex C}
\label{sec:dis_young_WR}

We found a young stellar population of $\sim7$\,Myr at the SN explosion site (position C).
\citet{2015MNRAS.451L..65T} found a very young stellar population of $\sim$ 3\,Myr at the SN site of PTF12dam, and furthermore linked this to a quantitative estimate of 60\,\msun for the SN progenitor. However, these data and analysis do not directly prove that the progenitor was of this age and mass. As discussed in \cite{2015MNRAS.452.1567C}, the bulk of the stellar population in the host of PTF12dam spans ages between 10 and 40\,Myrs. There is certainly evidence of a younger population of massive stars (a few Myrs old) existing from the emission lines, but this does not indicate that the progenitor was one of them. The same argument should apply to the host system of LSQ14mo.
The spatial resolution of the spectra and images are seeing dominated and 1\arcsec\ corresponds to 4\,kpc at $z = 0.256$. Therefore a direct age estimate is not reliable from ground-based spectra, and a quantitative mass is not supported as a unique turn off mass to apply for the progenitor mass of the SLSN.

The very deep spectrum of the host system of LSQ14mo allows us to search for Wolf-Rayet (WR) star features, which are a good tracer of recent massive star formation \citep{1981A&A...101L...5K}. There are two main WR features \citep[see e.g.][]{2008A&A...485..657B} seen in optical spectra: the blue (4600-4680 \AA) and red (5650-5800 \AA) bumps. 
However, we do not find these features in our spectra for components A, B, or C. 
We followed the method used in \citet{2015MNRAS.451L..65T} to determine the upper limit of WR star populations.
We measured the $3\sigma$ detection limit of \Heii $\lambda4686$ and $\ion{C}{IV}$ $\lambda5808$ lines assuming a full width at half maximum (FWHM) between 1000 and 3000 km\,s$^{-1}$ \citep{2008A&A...485..657B}. The luminosity of \Heii line is $<9.29\times10^{38}$ - $2.80\times10^{39}$\,erg\,s$^{-1}$, and that of $\ion{C}{IV}$ line is $<1.29\times10^{39}$ - $3.86\times10^{39}$\,erg\,s$^{-1}$. We also considered the metallicity-luminosity relation for WR stars using Eqs. 7 and 8 in \citet{2010A&A...516A.104L} and the host metallicity of $12 + \log({\rm O/H}) \sim 8.2$ (i.e. 0.3 solar). We obtained the upper limit of < 760-2280 of late-type WN stars, and < 670-2000 of early-type WC stars.

The lack of WR star features may indicate that either there is a very young stellar population environment, which is younger than 2\,Myr before WR stars evolved, or it may imply an older environment, such that WR stars have already died within 5\,Myr after a star-forming episode. The \ha EW indication of a 7\,Myr old stellar population at LSQ14mo host galaxy appears consistent with the latter explanation. 
In addition, there is no narrow galaxy \Heii $\lambda4686$ line detected in the entire host system of LSQ14mo. To date, only the host galaxy of PTF12dam has been reported to have this line detection. However, unambiguous WR features are not been detected
 \citep{2015MNRAS.452.1567C, 2015MNRAS.451L..65T}.

\section{Conclusions}
\label{sec:conclusions}

LSQ14mo is a spectroscopically normal type Ic SLSN at $z = 0.2563$, with a fast-declining lightcurve. We have presented its photometric and spectroscopic evolution from the early rising phase through to approximately +70\,d after maximum light. Combining our multicolour data and {\it K}-correction from spectra, we were able to build the bolometric lightcurve, which we fitted with magnetar and CSM interaction-powered models. As with other SLSNe, both scenarios can satisfactorily reproduce the observations. 
Using SN~1998bw as a template, we find that our upper limit of 2\msun $^{56}$Ni mass derived for LSQ14mo is not sufficient to power the peak luminosity of LSQ14mo, and thus we ruled out the purely $^{56}$Ni-powered mechanism. We fit our spectra of LSQ14mo with a synthetic spectral model \citep{2016MNRAS.458.3455M}. The early spectra can be reproduced assuming a magnetar-like energy source, but residuals from the later fits suggest that there may be a contribution from a partially thermalised, additional source of luminosity, starting after day $-7$\,d and ending before +22\,d. The spectrum at +9\,d is the most heavily affected, which shows a hot continuum but rather cool lines. We propose that this additional component could be produced by the interaction between the SN ejecta and a thin shell of material that was likely released during a pre-explosion mass-loss event of the massive progenitor star.

Late-time imaging of the SN revealed a host galaxy system with three resolved components. This was confirmed by deep VLT+FORS2 spectroscopy, which showed emission lines at three positions that are all spatially distinct and at slightly different velocities. LSQ14mo is coincident with the position we designated `C', which is the furthest from the brightest component in the system (designated `A'), at a velocity offset of $\approx 200$\,\kms. This is comparable to the relative velocities of the Magellanic Clouds relative to the Milky Way. The physical diameter of component C (A) was found to be 2.6\,kpc (6.2\,kpc), and stellar mass from the SED fitting was found to be $10^{7.7}$\,\msun ($10^{8.8}$\,\msun), which is also comparable to the Magellanic Clouds. By resolving the SN location in spectroscopy and imaging, we were able to carry out a detailed analysis comparing the SN site to the global properties of the host system. We found that the SN location had the lowest oxygen abundance of the three resolved components, with $12 + \log({\rm O/H}) = 7.92$-8.18, depending on the metallicity diagnostics used. This is $\sim$ 0.4 dex lower than the metallicity at position A. On the other hand, we measured a fairly constant specific SFR across the host system, and the specific SFR is typical compared to the general star-forming dwarf galaxies at similar redshift. 
Finally, we derive a young stellar population (6-9\,Myr) at all three positions as traced by the \ha line width. 
We propose that components C and A are an interacting galaxy system that has triggered star-formation, and that component B is a new-born \Hii region, which thus shows the strongest EW of \ha line and the weakest stellar continuum.  

The low metallicity constraint is favoured by the magnetar model as the power source for LSQ14mo. The spin-down period (3.9\,ms) and the host galaxy metallicity ($12 + \log{({\rm O/H})} = 8.18 \pm 0.05$, N2 scale) of LSQ14mo is in agreement with the magnetar spin-period and host metallicity relation proposed by \citet{2016arXiv160504925C}.

\section*{Acknowledgments}
T.-W.~Chen thanks the organisers and participants of MIAPP workshop `Supernovae: The Outliers' for stimulating discussions: You-Hua Chu for interacting galaxy, Maryam Modjaz for metallicity and Ken'ichi Nomoto for CSM model.
T.-W.~Chen thanks Stefan Taubenberger for bolometric lightcurve comparison and participants in the Garching SN meeting for useful advise. 
T.W.C, T.K., R.Y. and P.W. acknowledge the support through the Sofia Kovalevskaja Award to P. Schady from the Alexander von Humboldt Foundation of Germany. S.J.S. acknowledges ERC grant 291222 and STFC grants ST/I001123/1 and ST/L000709/1.
L.G. was supported in part by the US National Science Foundation under Grant AST-1311862.
A.J. acknowledges funding by the European Union's Framework Programme for Research and Innovation Horizon 2020 under Marie Sklodowska-Curie grant agreement No 702538.
M.S. acknowledges support from EU/FP7-ERC grant 615929.
K.M. acknowledges support from the STFC through an Ernest Rutherford Fellowship.

The Liverpool Telescope is operated on the island of La Palma by Liverpool John Moores University in the Spanish Observatorio del Roque de los Muchachos of the Instituto de Astrofisica de Canarias with financial support from the UK Science and Technology Facilities Council. Part of the funding for GROND (both hardware as well as personnel) was generously granted from the Leibniz-Prize to Prof. G. Hasinger (DFG grant HA 1850/28-1). 
We acknowlegde use of the Pan-STARRS1 photometric reference catalogue : the Pan-STARRS1 Surveys (PS1) have been made possible through contributions of the Institute for Astronomy, the University of Hawaii, the Pan-STARRS Project Office, the Max-Planck Society and its participating institutes, the Max Planck Institute for Astronomy, Heidelberg and the Max Planck Institute for Extraterrestrial Physics, Garching, The Johns Hopkins University, Durham University, the University of Edinburgh, Queen's University Belfast, the Harvard-Smithsonian Center for Astrophysics, the Las Cumbres Observatory Global Telescope Network Incorporated, the National Central University of Taiwan, the Space Telescope Science Institute, the National Aeronautics and Space Administration under Grant No. NNX08AR22G issued through the Planetary Science Division of the NASA Science Mission Directorate, the National Science Foundation under Grant No. AST-1238877, the University of Maryland, and Eotvos Lorand University (ELTE) and the Los Alamos National Laboratory.


\begin{appendix}

\section{Additional details of data reduction}
\label{sec:SN_photometry}

LSQ detected the SN during the early rising phase ($-7.8$ rest-frame days relative to the $r$-band maximum light on MJD = 56697).
We investigated pre-discovery LSQ images and found even earlier detections of this object at $-22.2$\,d. 
Although the LSQ pipeline reports magnitudes of the SN at each epoch at which it is detected, we opted to carry out our own point-spread function (PSF)-fitting photometry on every image in order to maintain consistency between epochs at which the pipeline did and did not detect SN flux automatically. 
The LSQ survey observations are carried out in a wide optical filter, effectively $g+r$. We calibrated our PSF photometry to $r$ band using a sequence of local field stars. We calculated the LSQ$-r$ colour on each night with a pipeline detection, and checked this against the expected LSQ$-r$ colour derived from synthetic photometry of the spectra of LSQ14mo, finding the two to be consistent. The calibrated photometry in the LSQ system was found to be very close to $r$ band, with LSQ$-r < 0.02$\,mag.

The LT images were reduced automatically by the LT pipeline, which applies de-bias and flat-field corrections. We extracted the SN flux using PSF-fitting, and calibrated the magnitudes using local sequence stars, which were themselves calibrated against standard fields observed with IO:O on photometric nights. 
We also obtained one epoch from Magellan + IMACS, reduced with bias subtraction and flat-fielding.
The NTT + EFOSC2 data were reduced using the PESSTO pipeline \citep{2015A&A...579A..40S} with bias subtraction and flat-fielding using twilight sky frames. Zeropoints were determined with nine Pan-STARRS1 catalogue sequence stars in the field \citep{2012ApJ...756..158S,2013ApJS..205...20M} and converted to SDSS photometric system \citep{2012ApJ...750...99T}, and PSF photometry was carried out 
using the {\sc daophot} task within {\sc iraf}. For the last two epochs of $g$ band from the NTT + EFOSC2, we have applied a bandpass correction of $+0.31$ mag from the $Gunn\,g$ to the SDSS $g$ \citep{2014ApJ...796...87I}.

After +250\,d, PESSTO took deep images to search for any late 
SN signal \citep[e.g. as in][]{2013ApJ...770..128I}.  
We used $i$-band data taken at +339.3\,d and $r$-band data taken at +321.7\,d as templates for image subtraction from the late epochs. The $R$-filter was misused for two epochs, and thus we used $R$-band data taken at +339.3\,d as a template to subtract the $R$-band image taken at +252.6\,d. 
We aligned the images using the {\sc geomap} and {\sc geotran} tasks within {\sc iraf} and employed the High Order Transform of PSF ANd Template Subtraction ({\sc hotpants})\footnote{http://www.astro.washington.edu/users/becker/v2.0/hotpants.html} package to carry out the subtraction. 
We did not detect a point source at the SN position from +252.6\,d to +321.7\,d, which indicates that either there is no SN signal or the SN does not change its brightness during these epochs. We assumed that the templates do not contain SN signal, as the measured host magnitude remains stable over these late epochs and no SN features are detected in deep ($\sim$5 hours of VLT exposure) VLT spectra obtained during the same phase (see Sect.\,\ref{sec:host_spectroscopy}). We therefore measured $3\sigma$ limiting magnitudes on the late-time images, by adding ten artificial stars at $\sim$ 0.5\arcmin\ around the SN position, and re-measuring their magnitudes until we obtained a standard deviation of 0.3\,mag.

We obtained pure host images on 2015 December 17 (+539.2\,d) using the Gamma-Ray Burst Optical/Near-Infrared Detector \citep[GROND;][]{2008PASP..120..405G}, a 7-channel imager that collects multi-band photometry simultaneously with $g'r'i'z'JHK$ bands, mounted at the 2.2 m MPG telescope at ESO La Silla Observatory in Chile. 
We used different aperture sizes to measure each component in each band.
The $griz$ photometry was calibrated against Pan-STARRS1 sequence stars and converted to the SDSS photometric system; while $JHK$ was calibrated against the 2MASS catalogue and converted to the AB system.
We noticed that the $g'$-band image has a higher light dispersion and the FWHM of stars are larger than other filters. This affected our flux measurement of the host components since they are not resolved clearly.
We compared GROND photometry to the host magnitude measurement from the NTT + EFOSC $i$-band image, because different seeing conditions cause differences in the PSF, and thus different aperture sizes were needed. The two datasets are in agreement with a systematic uncertainty of approximately 0.1\,mag.

For our host spectra, FORS2, with the standard resolution collimator, provides a pixel scale of 0.25\arcsec/pixel. 
We used an aperture size of 1.25\arcsec\ for trace A and a smaller aperture of 0.5\arcsec\ for the spatially-close traces B and C to avoid including fluxes from one to the other. Traces A and C are clearly identified from the emission lines and the stellar continuum. No continuum trace is visible at position B, but the nebular emission lines are clear; we therefore used the continuum trace of C as a reference to define the trace for position B. 
We employed a module of {\sc pysynphot} from {\sc python} to calculate synthetic photometric quantities for the extracted, flux-calibrated spectra A, B, and C in the NTT + EFOSC2 $i$-band bandpass, and then scaled the fluxes of spectra to match their $i$-band photometry (Sect.\,\ref{sec:host_photometry}), by a scaling factor of 2.72, 1.46, and 1.86 for spectra A, B and C, respectively.

\section{Emission line flux measurement}
\label{sec:line_measurements}

The spectrum A has strong stellar absorptions superimposed on the Balmer lines. Spectra B and C suffer from the same effect, but to a lesser extent (see Fig.\,\ref{fig:host_model_spec}). Hence, we employed the spectral synthesis code {\sc starlight} \citep{2005MNRAS.358..363C, 2009RMxAC..35..127C} to fit our observed spectra. 
We used the SSP model base of \citet{2007ASPC..374..303B} and selected 66 models from it. This selection base has been applied in \citet{2012A&A...545A..58S, 2014A&A...572A..38G, 2016MNRAS.455.4087G}, which consists of 66 components with 17 different ages (from 1\,Myr to 18\,Gyr) and four metallicities (0.2, 0.4, 1.0 and 2.5 \zsol, where \zsol = 0.02) coming from a slightly modified version of the models of \citet{2003MNRAS.344.1000B} and uses a Chabrier (2003) IMF. In general, the model fittings are good for three components with a $\chi^{2}$ of 2.22, 0.03, and 0.08 for positions A, B, and C, respectively.
The modelled spectra A, B, and C were subtracted from our observations and thus recover the true Balmer emission line fluxes (Fig.\,\ref{fig:host_model_spec}).

The spectrum A has the highest signal-to-noise ratio, and the main optical emission lines were all clearly detected; especially an intrinsic weak auroral \Oiii $\lambda4363$ line is marginally detected at $3.6 \sigma$. The spectrum B has the weakest continuum but the largest emission line equivalent widths. The spectrum C, where the SN is located, shows no sign of a broad \Oi $\lambda6300$ line, which would generally be the strongest type Ic SN feature in the nebular phase. Therefore, there is no evidence for supernova flux from LSQ14mo at this epoch. No \Heii $\lambda4686$ line is detected at any position in the host system. 

Line flux measurements were made after subtraction of the {\sc starlight} synthesis models. 
We fitted Gaussian line profiles using the QUB custom built {\sc procspec} environment within {\sc idl}. The FWHM for each line in a given spectrum was fixed to that of the \Oiii $\lambda4959$ line. The line width is fairly closely matched with the instrumental resolution ($\sim$ 9\AA\ in the observed-frame) measured from the skylines.
We further normalised the spectrum and determined the equivalent widths (EWs) of all identified lines. Uncertainties were estimated from the EW and the root mean square (rms) of the continuum according to the equation from \citet{1994ApJ...437..239G}:
\begin{equation}
\label{eq:flux_error}
 \sigma _{i} = \sigma _{c} \sqrt{N+\frac{W _{i}}{\Delta }}
,\end{equation}
where 
$\sigma _{i}$ is the error on the flux of the emission line, $\sigma _{c}$ is the {\it rms} measured from the local continuum, {\it N} is the measured line profile width in pixels, {\it $W _{i}$} is the absolute value of the line EW in  \AA, and {\it $\Delta$} is the spectral dispersion in \AA/pixel.
The flux measurements and related parameters are listed in Table\,\ref{tab:galaxy_flux}.

\section{Observations log}
\label{sec:app_log}

\begin{table*}
 \begin{minipage}{190mm}
 \centering
  \caption{Log of optical imaging of LSQ14mo. The phase (day) has been corrected for time dilation ($z = 0.256$) and is relative to the SN $r$-band maximum on MJD 56697. The `$>$' denotes the 3$\sigma$ detection limit. Optical $gri$ magnitudes are in the AB system and parenthesized with statistical errors. Note: $R$-band magnitudes are marked with a $^{*}$ symbol.   }
\label{tab:sn_phot_opi}
\begin{tabular}[t]{llllllllllll}
\hline\hline 
Date & MJD & Phase & $g$ (error) & $r$ (error)& $i$ (error) & Telescope & Instrument \\
\hline
2014 Jan 02     &56659.23       &$-30.1$        &               -       & $>$ 20.94       &-                      &ESO 1.0-m Schmidt& QUEST\\
2014 Jan 04     &56661.22       &$-28.5$        &               -       & $>$ 21.43       &-                      &ESO 1.0-m Schmidt& QUEST\\
2014 Jan 06     &56663.16       &$-26.9$        &       -               & $>$ 20.28       &-                      &ESO 1.0-m Schmidt& QUEST\\
2014 Jan 08     &56665.16       &$-25.4$        &       -               & $>$ 20.36       &-                      &ESO 1.0-m Schmidt& QUEST\\
2014 Jan 10     &56667.16       &$-23.8$        &       -               & $>$ 20.79       &-                      &ESO 1.0-m Schmidt& QUEST\\
2014 Jan 12     &56669.15       &$-22.2$        &               -       &20.82 (0.22)  &-              &ESO 1.0-m Schmidt& QUEST\\
2014 Jan 14     &56671.14       &$-20.5$        &       -               &21.04 (0.29)  &       -       &ESO 1.0-m Schmidt& QUEST\\
2014 Jan 16     &56673.14       &$-19.0$        &               -       &20.64 (0.22)  &       -       &ESO 1.0-m Schmidt& QUEST\\
2014 Jan 18     &56675.13       &$-17.4$        &               -       &20.41 (0.26)  &       -       &ESO 1.0-m Schmidt& QUEST\\
2014 Jan 20     &56677.12       &$-15.8$        &       -               &19.95 (0.20)  &       -       &ESO 1.0-m Schmidt& QUEST\\
2014 Jan 22     &56679.12       &$-14.2$        &               -       &19.99 (0.12)  &       -       &ESO 1.0-m Schmidt& QUEST\\
2014 Jan 24     &56681.14       &$-12.6$        &               -       &19.77 (0.21)  &       -       &ESO 1.0-m Schmidt& QUEST\\
2014 Jan 28     &56685.12       &$-9.4$ &               -       &19.92 (0.16)   &                 -&ESO 1.0-m Schmidt& QUEST\\
2014 Jan 30     &56687.17       &$-7.8$ &               -       &19.49 (0.14)   &                 -&ESO 1.0-m Schmidt& QUEST\\
2014 Feb 1      &56689.09       &$-6.3$ &       -               &19.40 (0.24)   &                 -&ESO 1.0-m Schmidt& QUEST\\
2014 Feb 3      &56691.09       &$-4.7$ &               -       &19.49 (0.11)   &                 -&ESO 1.0-m Schmidt& QUEST\\
2014 Feb 9      &56697.01       &0.0            &19.48 (0.04)   &19.42 (0.05)   &19.57 (0.06)  &LT        &IO\\        
2014 Feb 9      &56697.07       &0.1            &        -              &19.42 (0.19)  &               -&ESO 1.0-m Schmidt& QUEST\\
2014 Feb 10     &56698.03       &0.8            &19.52 (0.03)&19.50 (0.07)      &19.53 (0.07)  &LT        &IO\\        
2014 Feb 10     &56698.98       &1.6            &19.54 (0.04)&19.42 (0.05)      &19.49 (0.06)  &LT        &IO\\
2014 Feb 12     &56700.01       &2.4            &19.52 (0.06)&19.50 (0.10)      &19.50 (0.12)  &LT     &IO\\
2014 Feb 23     &56711.07       &11.2   &20.13 (0.07)&19.74 (0.08)      &19.74 (0.09)  &LT     &IO\\
2014 Feb 26     &56714.07       &13.6   &        -      &19.78 (0.16)   &         -       &ESO 1.0-m Schmidt& QUEST\\
2014 Feb 28     &56716.08       &15.2   &       - &19.86 (0.15) &       -               &ESO 1.0-m Schmidt& QUEST\\
2014 Mar 02     &56718.13       &16.8   &       -       &19.85 (0.21)   &         -       &ESO 1.0-m Schmidt& QUEST\\
2014 Mar 04     &56720.03       &18.3   &       -       &20.03 (0.19)   &         -       &ESO 1.0-m Schmidt& QUEST\\
2014 Mar 06     &56722.06       &20.0   &       -       &20.32 (0.15)   &         -       &ESO 1.0-m Schmidt& QUEST\\
2014 Mar 07     &56723.91       &21.4   &21.27 (0.20)&20.44 (0.07)      &20.11 (0.11)  &LT     &IO\\
2014 Mar 08     &56724.06       &21.7   & -&20.46 (0.16)        &       -               &ESO 1.0-m Schmidt& QUEST\\
2014 Mar 10     &56726.14       &23.2   &       -       &20.65 (0.20)   &                 -       &ESO 1.0-m Schmidt& QUEST\\
2014 Mar 12     &56728.05       &24.7   & -             &20.57 (0.19)   &                 -       &ESO 1.0-m Schmidt& QUEST\\
2014 Mar 12     &56728.91       &25.4   &21.48 (0.27)&20.80 (0.17)      &20.41 (0.13)  &LT     &IO\\
2014 Mar 18     &56734.13       &29.4   &       -       &21.21 (0.35)   &         -               &ESO 1.0-m Schmidt& QUEST\\
2014 Mar 20     &56736.88       &31.8   &21.78 (0.20)&20.97 (0.24)      &20.54 (0.16)  &LT     &IO\\
2014 Mar 24     &56740.23       &34.4   &       -       &21.11 (0.10)   &       -       &Magellan& IMACS\\
2014 Apr 21     &56768.02       &56.5   &23.77 (0.10)&22.42 (0.13)      &22.07 (0.16)  &NTT & EFOSC2\\
2014 May 6      &56783.01       &68.5   &24.14 (0.14)&22.85 (0.15)      &22.32 (0.13)  & NTT & EFOSC2\\
2014 Dec 23     &57014.31       &252.6  &               -       & $>$ 23.92 & $>$ 24.46& NTT & EFOSC2\\                     
2014 Dec 23             &57014.31       &252.6          &-      &  $>$ 23.79$^{*}$      &-      & NTT     &       EFOSC2 \\                       
2015 Jan 21     &57043.31       &275.7  &       -               &  $>$ 23.96    & $>$ 24.04        &NTT & EFOSC2 \\
2015 Feb 12     &57065.18       &293.1  &       -               &  $>$ 24.03    & $>$ 24.03        &NTT & EFOSC2 \\
2015 Mar 20     & 57101.07      &321.7  &       -               &  $>$ 23.63    & $>$ 23.34        &NTT & EFOSC2 \\
2015 Apr 11     & 57123.18      &339.3  &       -               &  $>$ 23.22$^{*}$      & $>$ 23.24        &NTT & EFOSC2 \\
\hline 
\end{tabular}
\end{minipage}
\end{table*}

\begin{table*}
 \begin{minipage}{190mm}
 \centering
  \caption{Ultraviolet photometry of LSQ14mo in the {\it Swift} UVOT bands.   The phase (day) has been corrected for time dilation ($z = 0.256$) and is relative to the SN {\it r}-band maximum on MJD 56697. {\it Swift} magnitudes are in the AB system, with the uncertainty including statistical and systematic errors.  }
\label{tab:sn_phot_uv}
\begin{tabular}[t]{lllllllll}
\hline\hline 
Date & MJD & Phase & $uvw2$ & $uvm2$ & $uvw1$ & $u$ \\
\hline
2014 Jan 31     &56688.76       &$-6.6$ &       -               & 21.49 (0.17)  &                       20.80 (0.16)  &       -                \\
2014 Feb 02     &56690.03       &$-5.5$ &21.81 (0.26)   &       -               &
                -       &19.62 (0.17)   \\
2014 Feb 04     &56692.03       &$-4.0$ &               -       &               -       &-
                        &19.41 (0.15)     \\
2014 Feb 05     &56693.05       &$-3.1$ &               -       &               -       &-
                        &19.92 (0.21)     \\            
2014 Feb 09     &56697.97       &0.8    &               -       &               -       &-
                        &19.96 (0.20)     \\
2014 Feb 11     &56699.72       &2.2    &               -       &       -               &-
                        &19.94 (0.20)     \\ 
\hline 
\end{tabular}
\end{minipage}
\end{table*}

\begin{table}
\caption{Emission line measurements of the host galaxy system of LSQ14mo in the observed frame. The spectra have been corrected for Milky Way reddening corresponding to a dust extinction of $A_{V} = 0.20$. 
We measured the emission line fluxes after subtracting the stellar continuum from the {\sc starlight} spectral synthesis models. The error is derived by the equation\,\ref{eq:flux_error}.
We fixed the FWHM widths at the values derived from the \Oiii $\lambda$4959 line, which correspond to the FORS2 resolution (9.05, 9.06, and 9.21\AA\ for positions A, B, and C, respectively). The relativity large rms and error of \Oiii $\lambda$5007 is due to a strong sky line nearby. 
For luminosity, we have not applied host galaxy dust reddening corrections. The internal host extinction at position A is $A_{V} = 0.26$ derived by the intrinsic ratio of \ha and \hb, and negligible at positions B and C. The unit of Flux, Error, RMS of $\times 10^{-18}$ erg s$^{-1}$ cm$^{-2}$, Luminosity  of $\times 10^{39}$ erg s$^{-1}$. }
\label{tab:galaxy_flux}
\begin{tabular}[htbp]{ccccccccccccc}
\hline\hline 
\multicolumn{7}{c}{{\bf Position A}} \\
\hline
Line & $\lambda$ (\AA) & Flux & Error &  RMS & EW (\AA) & Luminosity \\
\hline
\Oii &3727              &498            &3      &0.7    &58     &101\\
\Neiii &3868    &39.2   &3.5    &1.2    &4.6    &7.9\\
H$\delta$ &4102 &23.1   &2.3    &0.8    &2.7    &4.7\\
H$\gamma$ &4340 &53             &2      &0.6    &3      &11\\
\Oiii &4363     &4.63           &1.28   &0.47   &1.1    &0.94\\
H$\beta$ &4861  &137            &1      &0.4    &11     &28\\
\Oiii &4959     &121            &1      &0.3    &12     &25\\
\Oiii &5007     &332            &8      &2.1    &32     &67\\
\Oi &6300               &35             &2      &0.6    &5      &7\\
\Nii &6548              &11.6   &1.6    &0.6    &1.7    &2.4\\
H$\alpha$ &6563 &425            &3      &0.6    &68     &86\\
\Nii &6583              &48             &2      &0.6    &8      &10\\
\Sii &6717              &93             &3      &0.9    &15     &19\\
\Sii &6731              &60             &3      &0.9    &10     &12\\
\hline 
\multicolumn{7}{c}{{\bf Position B}} \\
\hline
Line & $\lambda$ (\AA) & Flux & Error &  RMS & EW (\AA)& Luminosity \\
\hline
\Oii &3727              &74     &0.8    &0.13   &90     &15\\
\Neiii &3868    &12     &0.4    &0.13   &9      &2.4\\
H$\delta$ &4102 &7.4    &0.33   &0.11   &4      &1.5\\
H$\gamma$ &4340 &13     &0.2    &0.06   &14     &2.7\\
\Oiii &4363     & -     & -     &-              &-      & -\\
H$\beta$ &4861  &31     &0.4    &0.08   &40     &6.3\\
\Oiii &4959     &39     &0.3    &0.08   &45     &8.0\\
\Oiii &5007     &119&3.5        &0.49   &144    &24\\
\Oi &6300               &6.1&0.63&0.20  &11     &1.2\\
\Nii &6548              & -     & -     &-              &-      & -\\
H$\alpha$ &6563 &80     &1.1    &0.13   &219    &16\\
\Nii &6583              &2.7&0.41&0.13  &9      &0.6\\
\Sii &6717              &7.8&0.79&0.21  &23     &1.6\\
\Sii &6731              &6.9&0.81&0.21  &25     &1.4\\
\hline 
\multicolumn{7}{c}{{\bf Position C}} \\
\hline
Line & $\lambda$ (\AA)& Flux & Error &  RMS & EW (\AA)& Luminosity \\
\hline
\Oii &3727              &53     &1.0            &0.19   &60     &10.8\\
\Neiii &3868    &8.8    &0.54           &0.15   &9      &1.8\\
H$\delta$ &4102 & -     & -             &-       -      & -&\\
H$\gamma$ &4340 &9.3    &0.43           &0.13   &4      &1.9\\
\Oiii &4363     & -     & -             &-       -      & -&\\
H$\beta$ &4861  &23     &0.5            &0.12   &21     &4.8\\
\Oiii &4959     &26     &0.7            &0.18   &23     &5.2\\
\Oiii &5007     &77     &6.2            &1.11   &70     &15.7\\
\Oi &6300               & -     & -             &-       -      & -&\\
\Nii &6548              & -     & -             &-       -      & -&\\
H$\alpha$ &6563 &64     &1.5            &0.24   &107&13.0\\
\Nii &6583              &3.7&0.83       &0.24   &8      &0.7\\
\Sii &6717              &6.8&0.81       &0.22   &13     &1.4\\
\Sii &6731              &5.0&0.80       &0.22   &11     &1.0\\
\hline 
\end{tabular}
\end{table}

\begin{table}
 \begin{minipage}{83mm}
 \centering
  \caption{{\it K}-corrections for LSQ14mo and its host galaxy system derived from the spectroscopic observations using {\sc snake} \citep{2016arXiv160401226I}.  The phase (day) has been corrected for time dilation ($z = 0.256$) and is relative to the SN $r$-band maximum on MJD 56697. The MJD and phase of the host galaxy was adopted from the middle time of three VLT observations.}
\label{tab:sn_K-correction}
\begin{tabular}[t]{llllllllll}
\hline\hline 
Date & MJD & Phase & $K_{g \rightarrow u}$ & $K_{r \rightarrow g}$ & $K_{i \rightarrow r}$ \\
\hline
LSQ14mo \\
2014 Jan 31     & 56688.16      & -7.0          & 0.20 & 0.26 & 0.29 \\
2014 Feb 6      & 56694.19      & -2.2          & 0.22 & 0.24 & 0.28 \\
2014 Feb 20     & 56708.06      & 8.8   & 0.28 & 0.27 & 0.26 \\
2014 Feb 28     & 56716.13      & 15.2  & 0.37 & 0.34 & 0.26 \\
2014 Mar 8      & 56724.26      & 21.7  & 0.41 & 0.38 & 0.25 \\
2014 Apr 21     & 56768.04      & 56.5  & 0.38 & 0.32 & 0.31 \\
\hline
Host galaxy &    &   &   &   &   \\
2015 Jan-Feb  & 57065.58 & 293.5 &   &   &   \\
A &   &   &  0.57  &  0.33  &  0.21  \\
B&    &  &  0.44  &  0.35  &  0.15 \\
C&    &   &  0.58 &  0.37 &  0.19 \\
\hline
\end{tabular}
\end{minipage}
\end{table}

\begin{table}
 \begin{minipage}{83mm}
 \centering
  \caption{Final and optical pseudo bolometric luminosities of LSQ14mo with their uncertainties. The SN phase is reported with respect to the time of observed maximum $r$-band light in the rest-frame. Note: The epoch of bolometric lightcurve corresponds to photometric epoch, except those epochs $-7.0$, $-2.2$ and +8.8, which were derived from the spectroscopy, and epochs 0.1 and 29.4 were dropped due to large photometry errors.} 
\label{tab:bol_flux}
\begin{tabular}{c c c}
\hline\hline 
Phase & $\mathit{log}\,{\rm L_{bol}}$ & $\mathit{log}\,{\rm L_{pseudo}}$\\  
(day) & (erg/s)  & (erg/s)  \\  
\hline
-22.2&  $43.47  \pm 0.05$&$43.16\pm0.04$\\
-20.5&  $43.38  \pm 0.05$&$43.08\pm0.04$\\
-19.0&  $43.55  \pm 0.04$&$43.24\pm0.04$\\
-17.4&  $43.64  \pm 0.04$&$43.33\pm0.04$\\
-15.8&  $43.83  \pm 0.04$&$43.52\pm0.03$\\
-14.2&  $43.81  \pm 0.04$&$43.50\pm0.03$\\
-12.6&  $43.90  \pm 0.04$&$43.59\pm0.03$\\
-9.4    &$43.84 \pm 0.04$&$43.54\pm0.03$\\
-7.8    &$44.01 \pm 0.04$&$43.71\pm0.03$\\
-7.0    &$44.03 \pm 0.03$&$43.73\pm0.03$\\
-6.3    &$44.04 \pm 0.03$&$43.75\pm0.03$\\
-4.7    &$44.03 \pm 0.03$&$43.74\pm0.03$\\
-2.2    &$44.02 \pm 0.03$&$43.74\pm0.03$\\
0.0     &$43.99 \pm 0.03$&$43.72\pm0.02$\\
0.8     &$43.99 \pm 0.03$&$43.72\pm0.02$\\
1.6     &$43.98 \pm 0.03$&$43.73\pm0.02$\\
2.4     &$43.97 \pm 0.02$&$43.72\pm0.02$\\
8.8     &$43.86 \pm 0.03$&$43.64\pm0.03$\\
11.2    &$43.79 \pm 0.02$&$43.61\pm0.02$\\
13.6    &$43.74 \pm 0.02$&$43.58\pm0.03$\\
15.2    &$43.69 \pm 0.02$&$43.56\pm0.04$\\
16.8    &$43.67 \pm 0.02$&$43.55\pm0.03$\\
18.3    &$43.62 \pm 0.02$&$43.50\pm0.04$\\
20.0    &$43.57 \pm 0.02$&$43.44\pm0.04$\\
21.4    &$43.52 \pm 0.02$&$43.39\pm0.04$\\
21.7    &$43.52 \pm 0.02$&$43.38\pm0.04$\\
23.2    &$43.46 \pm 0.02$&$43.34\pm0.04$\\
25.4    &$43.39 \pm 0.03$&$43.27\pm0.05$\\
31.8    &$43.28 \pm 0.02$&$43.17\pm0.05$\\
34.4    &$43.19 \pm 0.02$&$43.07\pm0.05$\\
56.5    &$42.83 \pm 0.03$&$42.60\pm0.05$\\
68.5    &$42.67 \pm 0.04$&$42.41\pm0.04$\\
252.6   &$<41.86$&$ <41.44$\\ 
275.7   &$<41.88$&$<41.47$\\ 
293.1   &$<41.85$&$<41.45$\\ 
321.7   &$<42.04$&$<41.65$\\ 
\hline 
\end{tabular}
\end{minipage}
\end{table}

\end{appendix}
%
%



\end{document}